\documentclass[aps,prb,twocolumn,superscriptaddress,showpacs]{revtex4-1}
\usepackage{amsmath,amssymb,graphics,epsfig,epstopdf,color,verbatim,ulem,braket,tabularx}
\usepackage[colorlinks,linkcolor=blue,citecolor=blue,urlcolor=blue]{hyperref}

\newcommand{\COMMENTED}[1]{}

\begin{document}

\title{Finite-temperature Auxiliary-Field Quantum Monte Carlo: \\
Self-Consistent Constraint and Systematic Approach to Low Temperatures}

\author{Yuan-Yao He}
\address{Department of Physics, Renmin University of China, Beijing 100872, China}
\address{Department of Physics, College of William and Mary, Williamsburg, Virginia 23187, USA}
\address{Center for Computational Quantum Physics, Flatiron Institute, New York, New York 10010, USA}
\author{Mingpu Qin}
\address{Department of Physics, College of William and Mary, Williamsburg, Virginia 23187, USA}
\author{Hao Shi}
\address{Center for Computational Quantum Physics, Flatiron Institute, New York, New York 10010, USA}
\author{Zhong-Yi Lu}
\address{Department of Physics, Renmin University of China, Beijing 100872, China}
\author{Shiwei Zhang}
\address{Department of Physics, College of William and Mary, Williamsburg, Virginia 23187, USA}
\address{Center for Computational Quantum Physics, Flatiron Institute, New York, New York 10010, USA}

\begin{abstract}
We describe an approach for many-body calculations with a  finite-temperature, grand canonical ensemble formalism using auxiliary-field quantum Monte Carlo (AFQMC) with a self-consistent constraint to control the sign problem. The usual AFQMC formalism of Blankenbecler, Scalapino, and Sugar suffers from the sign problem with most physical Hamiltonians, as is well known. Building on earlier ideas to constrain the paths in auxiliary-field space [Phys. Rev. Lett. \textbf{83}, 2777 (1999)] and incorporating recent developments in zero-temperature, canonical-ensemble methods, we discuss how a self-consistent constraint can be introduced in the finite-temperature, grand-canonical-ensemble framework. This together with several other algorithmic improvements discussed here leads to a more accurate, more efficient, and numerically more stable approach for finite-temperature calculations. We carry out a systematic benchmark study in the two-dimensional repulsive Hubbard model at $1/8$ doping. Temperatures as low as $T=1/80$ (in units of hopping) are reached. The finite-temperature method is exact at very high temperatures, and approaches the result of the zero-temperature constrained-path AFQMC as temperature is lowered. The benchmark shows that  systematically accurate results are obtained for thermodynamic properties.
\end{abstract}


\date{\today} \maketitle

\section{Introduction}
\label{sec:Introduction}

Quantum Monte Carlo (QMC) methods have become a key numerical technique for studying interacting quantum many-body systems in various areas, including condensed matter~\cite{Blankenbecler1981,Sugiyama1986}, high energy~\cite{Hands1993,Chandrasekharan2010} and nuclear~\cite{Carlson2015} physics as well as quantum chemistry~\cite{Hammond1994}. Among them, auxiliary-field Quantum Monte Carlo (AFQMC) methods~\cite{Blankenbecler1981,Scalapino1981,Sugiyama1986,White1989,Shiwei1995,Shiwei1999,AssaadEvertz2008,Shiwei2013Review} decouple the two-body interactions in the Hamiltonian by the Hubbard-Stratonovich transformation~\cite{Hubbard1959}, and sample the resulting external auxiliary fields by Monte Carlo. AFQMC methods can produce accurate, sometimes even numerically exact, solutions in correlated fermion systems, by explicitly accessing only a small fraction of the whole auxiliary-field space, whose size grows exponentially with system size.

The determinantal quantum Monte Carlo (DQMC) algorithm, first formulated by Blankenbecler, Scalapino, and Sugar~\cite{Blankenbecler1981})(BSS), is commonly applied for both finite-temperature and ground-state~\cite{Sugiyama1986,sorella2017} calculations. Direct DQMC calculations, while formally exact, suffer from the minus sign problem~\cite{Schmidt1984,Loh1990,Santos2003} in general. As a result, the computational cost of DQMC calculations, for fixed statistical accuracy, scales exponentially with system size and inverse temperature~\cite{Loh1990,Santos2003}, instead of the polynomial scaling expected for systems free of the sign problem. Though many specific models have turned out to be free of the sign problem~\cite{Hirsch1985,Congjun2005,Berg2012,Huffman2014,WangLei2014,Zixiang2015,Wang2015,Zixiang2016,ZhongchaoWei2016,Huffman2016}, the vast majority of correlated electron systems are not. The sign problem hinders or prevents DQMC studies of a variety interesting problems in correlated fermion systems, including in Hubbard-like models with doping and repulsive interactions, and almost all realistic Hamiltonians of molecules and solids.

More recently, constrained-path (CP) AFQMC methods~\cite{Shiwei1995,Shiwei1999,Shiwei2003} have been developed as an alternative approach. Building on the basic formalism of the DQMC algorithm, these methods introduce a computational framework of random walks in the manifold of mean-field or independent-particle solutions, which connects with concepts in electronic structure and quantum chemistry~\cite{Shiwei2013Review}. CP-AFQMC controls the sign problem by applying a constraint which restricts the Monte Carlo sampling in auxiliary-field space. The constraint identifies and removes redundant contributions to the ground-state wave function or the finite-temperature density matrix by paths in auxiliary-field space which would cancel with explicit, full path integration but appear as random noise in Monte Carlo sampling. The formalism is exact when the path identification is exact. In practice it is implemented approximately with a trial wave function or trial density matrix, which introduces a possible systematic bias in the numerical results but in turn removes the exponentially growing computational cost and recovers the algebraic complexity. Through many tests and developments in the last two decades~\cite{Shiwei1997,Zhang1997,Guerrero1998,Carlson1999,Enjalran2001,Chang2008,Chang2010,Shihao2013,
Shihao2014,LeBlanc2015,MingPu2016a,MingPu2016b,Zheng2017}, the zero-temperature (ZT) CP-AFQMC method has been proved to be a highly accurate, general numerical approach for studying ground-state properties of various interacting fermion models as well as molecules and realistic materials by its generalization, the phaseless AFQMC method~\cite{Shiwei2003,Suewattana2007,Motta2017}.

The finite-temperature (FT) CP-AFQMC method was formulated to study thermodynamic properties in correlated fermion systems in Ref.~\onlinecite{Shiwei1999}. The constraint applied in the FT-CP-AFQMC method to control the sign problem involves an input trial Hamiltonian or trial density matrix, corresponding to the trial wavefunction used in ZT method. The FT-CP-AFQMC formalism has been extended to Bose-Fermi mixtures~\cite{Rubenstein2012}, and more recently applied to molecular systems~\cite{Liu2018} by replacing the sign constraint with the phaseless constraint~\cite{Shiwei2003}. However, applications of the FT-CP-AFQMC have been limited, since the benchmark has not been nearly extensive as for ZT, and the many recent developments in understanding and improving the constraint have not been realized within the FT framework.

The ability to compute the temperature dependence of various physical properties in many fermion systems is of fundamental importance. Examples in correlated fermion models include pseudogap physics in cuprates, as well as BCS-BEC crossover for ultracold fermions in optical lattice. More recently, the direct measurements of the equation of state~\cite{Cocchi2016}, Mott insulator~\cite{Cheuk2016} as well as short-range charge and spin correlations~\cite{Boll1257,Cheuk1260} of Hubbard Model are realized in optical lattice. As experiments start to access lower temperatures, there is a growing need for accurate computations across temperature ranges. It is especially timely to revisit the difficulty posed by the sign problem and extend the reach of AFQMC to finite but sufficiently low temperatures.

We address this need in the present paper. Building on the FT-CP-AFQMC ideas in Ref.~\onlinecite{Shiwei1999} and recent developments in ZT methods~\cite{,MingPu2016b,LeBlanc2015,MingPu2016a}, we introduce a self-consistent constraint in the FT framework, investigate the systematic error and its relation to spin symmetry breaking in the trial density matrix, and perform systematic benchmark studies in the two-dimensional Hubbard model. We also describe in detail several other algorithmic improvements in the FT framework which allow stable and efficient computations to low temperatures (as low as $T=1/80$ in units of hopping in the Hubbard model).

In the benchmark, we mainly concentrate on $1/8$ hole doping in the one-band Hubbard model, which has the most severe sign problem and is algorithmically very difficult. Physically it is a crucial regime to understand in the context of the phase diagram of cuprates. The results demonstrate that FT-CP-AFQMC is exact at very high temperatures and approaches ZT-CP with decreasing temperature. They show that the new FT-CP algorithm is capable of computing thermodynamic properties systematically with accuracy comparable to or even better than that of ZT-CP for ground-state properties.

In the rest of the paper, we will refer to the standard algorithm of BSS as DQMC and, unless necessary, refer to the CP-AFQMC framework as AFQMC for brevity, distinguishing the zero- and finite-temperature formalisms by ZT and FT respectively. As we will see in more detail below, the CP-AFQMC framework shares the basic formalism of DQMC; however, it reformulates the Monte Carlo process as branching random walks, using importance-sampling transformation, with an ensemble or population of walkers in the imaginary-time (inverse temperature) direction instead of the usual Metropolis sampling of the entire path in auxiliary-field space. The reason for the reformulation is two-fold. One is to allow the imposition of the constraints without incurring ergodicity problems~\cite{FahyHamann1990,Shiwei1997,Shiwei1999}. The other is that, for ground-state or ZT calculations, it provides a clear and formal connection~\cite{Shiwei2003,Shiwei2018Review} with standard electronic structure methods within density-functional theory, which has enabled general computations in solids and molecular systems.

The rest of the paper is organized as follows. In Sec.~\ref{sec:DQMCMethod}, we summarize the DQMC algorithm, and establish the necessary concepts and formalisms which are also important parts of FT-AFQMC  method. In Sec.~\ref{sec:FTCPMCMethod}, we describe the FT-AFQMC method, focusing on the constrained path approximation, process of random walks with importance sampling and several implementation issues for improved efficiency and stability in the numerical procedures. Then the self-consistent constraint and the forms of the constraining trial Hamiltonian or density matrix are discussed in Sec.~\ref{sec:FTCPMCTralSelf}, using 2D doped Hubbard model with repulsive interaction as a concrete example. In Sec.~\ref{sec:NumericalResults}, we present the benchmark results for this system, which demonstrates the performance and accuracy of the FT-AFQMC method. Finally, Sec.~\ref{sec:SumDiscuss} summarizes this work, and discusses applications of the method in other models and realistic molecules and materials.

\section{Determinantal quantum Monte Carlo method and the sign problem}
\label{sec:DQMCMethod}

There are various overlapping numerical details in DQMC and FT-AFQMC methods. Here we present a brief review of the DQMC method including the basic formalism and the sign problem. This will help illustrate the connection, and also introduce several concepts and set up out notation for describing AFQMC in the following sections.

\subsection{The formalism of DQMC method}
\label{sec:DQMCBasics}

We begin with a general interacting model of spin-$1/2$ fermions, $\hat{H}=\hat{H}_0+\hat{H}_I$, where $\hat{H}_0$ is the non-interacting part and $\hat{H}_I$ represents fermion-fermion interactions. $\hat{H}_0$ can be written as $\sum_{ij,\sigma\sigma^{\prime}}(\mathbf{H}_0)_{i\sigma,j\sigma^{\prime}} c_{i\sigma}^+c_{j\sigma^{\prime}}$($\sigma=\uparrow,\downarrow$ is the spin index), with $c_{i\sigma}^+$ ($c_{i\sigma}$) as creation (annihilation) operator on site $i$. A chemical potential term is included in $\hat{H}_0$ implicitly. Given a finite-size lattice (or basis) with $N_s$ sites, $\mathbf{\mathbf{H}_0}=\{(\mathbf{H}_0)_{i\sigma,j\sigma^{\prime}}\}$ is $2N_s\times 2N_s$ hopping matrix. If there are no spin-flip terms, $\mathbf{H}_0$ is block-diagonal with respect to spin species: $\mathbf{H}_0=\text{Diag}(\mathbf{H}_0^{\uparrow},\mathbf{H}_0^{\downarrow})$, where $\mathbf{H}_0^{\sigma}$ is $N_s\times N_s$ matrix.

The DQMC method deals with the partition function of the system as
\begin{eqnarray}
\label{eq:PartitionFunc0}
Z=\text{Tr}(e^{-\beta\hat{H}}) = \text{Tr}(\underbrace{e^{-\Delta\tau\hat{H}}\cdots e^{-\Delta\tau\hat{H}}e^{-\Delta\tau\hat{H}}}_M),
\end{eqnarray}
where $\Delta\tau=\beta/M$ and $M$ is the number of imaginary-time slices. For a small $\Delta\tau$, the Trotter-Suzuki decomposition, such as the asymmetric one $e^{-\Delta\tau\hat{H}}=e^{-\Delta\tau\hat{H}_I}e^{-\Delta\tau\hat{H}_0} + \mathcal{O}[(\Delta\tau)^2]$, and the Hubbard-Stratonovich transformation, generally expressed as $e^{-\Delta\tau\hat{H}_I} = \sum_{\mathbf{x}} p(\mathbf{x})\hat{B}_I(\mathbf{x})$, are applied to transform the many-body propagator $e^{-\Delta\tau\hat{H}}$ in Eq.~(\ref{eq:PartitionFunc0}) into single-particle operators expressed as free fermions coupled to auxiliary fields $\mathbf{x}=(x_{1},x_{2},\cdots,x_{N_f})$ with $N_f$ (comparable with $N_s$) components. The error $\mathcal{O}[(\Delta\tau)^2]$ originates from $[\hat{H}_0,\hat{H}_I]\ne0$, and $\hat{B}_I(\mathbf{x}) = \exp\big\{ \sum_{ij,\sigma\sigma^{\prime}}[\mathbf{H}_I(\mathbf{x})]_{i\sigma,j\sigma^{\prime}} c_{i\sigma}^+c_{j\sigma^{\prime}}\big\}$ where $\mathbf{H}_I(\mathbf{x})=\{[\mathbf{H}_I(\mathbf{x})]_{i\sigma,j\sigma^{\prime}}\}$ is a $2N_s\times 2N_s$ Hermitian or anti-Hermitian matrix. Combining with the kinetic propagator $\hat{B}_K=e^{-\Delta\tau\hat{H}_0}$ which has no dependence on imaginary time, we can rewrite the propagator $e^{-\Delta\tau\hat{H}(\ell)}$ at the $\ell$-th time slice as
\begin{eqnarray}
\label{eq:HSTransform}
e^{-\Delta\tau\hat{H}}
= \sum_{\mathbf{x}_{\ell}}p(\mathbf{x}_{\ell})\hat{B}_{\ell} + \mathcal{O}[(\Delta\tau)^2],
\end{eqnarray}
where $\hat{B}_{\ell}=\hat{B}_I(\mathbf{x}_{\ell})\hat{B}_K$. Applying this to all time slices in Eq.~(\ref{eq:PartitionFunc0}), we arrive at $Z\simeq \sum_{\mathbf{X}}P(\mathbf{X})\text{Tr}( \hat{B}_M\cdots\hat{B}_2\hat{B}_1 )$, where $P(\mathbf{X})=\prod_{\ell=1}^{M}p(\mathbf{x}_{\ell})$ is a probability density function and the auxiliary-field configuration $\mathbf{X}=\{\mathbf{x}_M,\cdots,\mathbf{x}_2,\mathbf{x}_1\}$ contains $MN_f$ components. Since $\hat{B}_{\ell}$ is a single-particle propagator, the trace in the partition function can now be evaluated explicitly to yield
\begin{eqnarray}
\label{eq:PartitionFunc1}
Z \simeq \sum_{\mathbf{X}}P(\mathbf{X})\text{det}(\mathbf{I}_{2N_s}+\mathbf{B}_M\cdots\ \mathbf{B}_2\mathbf{B}_1),
\end{eqnarray}
where $\mathbf{B}_{\ell}=\mathbf{B}_I(\mathbf{x}_\ell)\mathbf{B}_K$ with $\mathbf{B}_I(\mathbf{x}_\ell)=e^{\mathbf{H}_I(\mathbf{x}_{\ell})}$ and $\mathbf{B}_K=e^{-\Delta\tau\mathbf{H}_0}$. In Eq.~(\ref{eq:PartitionFunc1}), $D(\mathbf{X})=P(\mathbf{X})\text{det}(\mathbf{I}_{2N_s}+\mathbf{B}_M\cdots\mathbf{B}_2\mathbf{B}_1)$ is the so-called "weight" of auxiliary field configuration $\mathbf{X}$. With this, we have formally mapped the study of a $D$ dimensional quantum systems for fermions with $N_s$ lattice sites into solving a $(D+1)$ classical systems with $MN_f$ classical variables (or sites on a space-time lattice).

If the determinant satisfies $D(\mathbf{X})\ge0$, we can define a probability density function
\begin{equation}
\label{eq:W_X}
W(\mathbf{X})=\frac{D(\mathbf{X})}{\sum_{\mathbf{X}^{\prime}}D(\mathbf{X}^{\prime})}\,
\end{equation}
and sample it by Monte Carlo (for example with the Metropolis algorithm) to calculate physical observables
\begin{equation}
\label{eq:PhyObs}
\langle\hat{O}\rangle=\frac{1}{Z}\text{Tr}\big[e^{-(\beta-\tau)\hat{H}}\hat{O}e^{-\tau\hat{H}}\big] = \sum_{\mathbf{X}}O(\mathbf{X})W(\mathbf{X}),
\end{equation}
where the first and second equalities give the definition and the formula used in Monte Carlo calculations, respectively. In Eq.~(\ref{eq:PhyObs}), the extra $\tau=\ell\Delta\tau$ with $0\le\ell\le M$ is used to indicate that, although there is overall translational invariance in $\tau$ for $\langle\hat{O}\rangle$,
the evaluation of $O(\mathbf{X})$, the measurement of $\hat{O}$ within configuration $\mathbf{X}$, can depend on imaginary time. For example, the static Green's function matrix $\mathbf{G}(\tau,\tau)=\{G_{i\sigma,j\sigma^\prime}=\langle c_{i\sigma}c_{j\sigma^\prime}^+\rangle_{\tau}\}$ takes the simple form $\mathbf{G}(\tau,\tau)=(\mathbf{I}_{2N_s}+\mathbf{R}\mathbf{L})^{-1}$ where $\mathbf{R}=\mathbf{B}_{\ell}\cdots\mathbf{B}_{2}\mathbf{B}_{1}$ and $\mathbf{L}=\mathbf{B}_{M}\mathbf{B}_{M-1}\cdots\mathbf{B}_{\ell+1}$. With this, the $O(\mathbf{X})$ for all single-particle and two-particle observables can be calculated directly or via Wick decomposition. Eq.~(\ref{eq:PhyObs}) is then used to obtain the statistical average and perform standard analysis to estimate the statistical error.

We remark on several technical aspects. (1) For the special case of spin decoupled systems, the matrices involved become block diagonal with respect to spin species, as mentioned earlier. The determinant $D(\mathbf{X})$ can be separated as $D(\mathbf{X})=D_{\uparrow}(\mathbf{X})D_{\downarrow}(\mathbf{X})$.
(2) We used the asymmetric Trotter-Suzuki decomposition in the discussion above. It is straightforward to apply the symmetric one, e.g., $e^{-\Delta\tau\hat{H}}=e^{-\Delta\tau\hat{H}_0/2}e^{-\Delta\tau\hat{H}_I}e^{-\Delta\tau\hat{H}_0/2} + \mathcal{O}[(\Delta\tau)^3]$. In any case, the systematic error from a finite $\Delta\tau$ can be removed by extrapolating several calculations with different $\Delta\tau$ values. (3) The DQMC approach is typically realized by a  Markov-Chain Monte Carlo algorithm. Ergodicity is required in the sampling of new auxiliary-field configurations and detailed balance must be maintained in the update procedure. (4) The computational complexity of the DQMC algorithm is $\mathcal{O}(M N_s^3)$. (5) There can be an infinite variance problem \cite{HaoShiwei2016} in sign-problem-free simulations with $D(\mathbf{X})\ge0$ which needs to be controlled. We will defer the discussion of numerical stablization and efficient updating of auxiliary-field configurations to the sections on AFQMC, and discuss the sign problem in DQMC next.

\subsection{The minus sign problem}
\label{sec:SignProblem}

For systems free of the sign or phase problem, special symmetries are typically present to keep $D(\mathbf{X})\ge0$ for all configurations, such as the anti-unitary symmetry for the one-particle Hamiltonian after HS transformation~\cite{Hirsch1985,Congjun2005}. Recently, the guiding principles for interacting fermion models which can be made free of the sign or phase problem have been largely extended~\cite{Berg2012,Huffman2014,WangLei2014,Zixiang2015,Wang2015,Zixiang2016,ZhongchaoWei2016,Huffman2016}. However, these cases are still rare. In general, the determinant $D(\mathbf{X})$ cannot be made non-negative for all configurations and negative for others, which leads to the minus sign problem. More generally, the $\mathbf{B}_{\ell}$ matrix can be complex, because of hopping terms in $ \mathbf{H}_0$ from a twist angle or magnetic field, or spin-orbit coupling, and/or because of specific choices of the Hubbard-Stratonovich transformation in $\mathbf{H}_I$. This leads to a phase problem.

In the presence of a sign or phase problem,  $W(\mathbf{X})$ in Eq.~(\ref{eq:W_X}) can no longer be used as a probability density. It is of course straightforward to choose, for example, $W^\prime(\mathbf{X})=\frac{|D(\mathbf{X})|}{\sum_{\mathbf{X}^{\prime}}|D(\mathbf{X}^{\prime})|}$
instead, and replace the Monte Carlo average of the observable $\hat{O}$ in Eq.~(\ref{eq:PhyObs}) with
\begin{equation}
\label{eq:PhyObsNew}
\langle\hat{O}\rangle = \frac{ \sum_{\mathbf{X}} \text{sgn}(\mathbf{X}) O(\mathbf{X}) W^{\prime}(\mathbf{X}) }
        { \sum_{\mathbf{X}^{\prime}} \text{sgn}(\mathbf{X}^{\prime}) W^{\prime}(\mathbf{X}^{\prime}) },
\end{equation}
where $\text{sgn}(\mathbf{X})=D(\mathbf{X})/|D(\mathbf{X})|$ is the phase or sign of $D(\mathbf{X})$. The denominator in Eq.~(\ref{eq:PhyObsNew}) gives the average of the sign (or phase):
\begin{equation}
\label{eq:SignAverage}
\langle\text{sgn}\rangle = \sum_{\mathbf{X}} \text{sgn}(\mathbf{X}) W^{\prime}(\mathbf{X})
= \frac{\sum_{\mathbf{X}}D(\mathbf{X})}
              {\sum_{\mathbf{X}^{\prime}}|D(\mathbf{X}^{\prime})|},
\end{equation}
which can be thought of as the ratio between two partition functions, one for our actual interacting fermion system and the other a fictitious systems defined by the absolute value of the determinants. Note that $\langle\text{sgn}\rangle$ is always a real number and below we will loosely refer to it vanishing as a sign problem, although in the case of a phase problem the imaginary part can grow to be of comparable magnitude as the real part, causing large fluctuations in the phase.

Implementing the above reweighting technique does not prevent the sign problem, of course. It can be shown both theoretically and numerically~\cite{Loh1990,Santos2003} that $\langle\text{sgn}\rangle\propto e^{-\xi \beta N_s}$ for $\beta$ larger than a specific value, where $\xi$ is a positive constant dependent on the filling and interaction strength of the system. Thus, as $\beta$ increases, $D(\mathbf{X})$ tends to approach an antisymmetric function and the sign average $\langle\text{sgn}\rangle$ vanishes exponentially. The computational cost of DQMC simulations, for fixed statistical accuracy, scales exponentially with system size and inverse temperature for any $\langle\hat{O}\rangle$. As pointed out in Ref.~\onlinecite{Shiwei1999} and further discussed below in Sec.~\ref{sec:FTCPMCApprox}, the symptoms and the origin of the sign problem can be understood with a simple intuitive picture in terms of the path integrals in auxiliary-field space.

\section{Finite-temperature constrained path auxiliary-field quantum Monte Carlo Method}
\label{sec:FTCPMCMethod}

The FT-AFQMC approach is an alternative method. A constraint is applied, based on exact considerations of the nature of the paths in auxiliary-field space but implemented approximately using a trial density matrix, to control the sign/phase in  $D(\mathbf{X})$. The approximation can introduce a systematic error, but removes completely any decay of $\langle\text{sgn}\rangle$, and restores algebraic scaling. In this section, we discuss the basic formalism of FT-AFQMC method in the first part. Then we introduce the formulation of branching random walk with importance sampling, which is necessary for an efficient implementation. The last part contains several important technical aspects for improving numerical efficiency and stability.

\subsection{The constrained path approximation}
\label{sec:FTCPMCApprox}

The CP approaches at both finite- and zero-temperatures are based on understanding the structure of the paths in auxiliary-field space, and devising rigorous constraints which effectively sum positive and negative contributions prior to the Monte Carlo sampling. Below we illustrate the idea at finite-temperature
using the formalism of DQMC discussed earlier.

Instead of starting from a full initial configuration $\mathbf{X}$ as in DQMC, we imagine that the complete path $\mathbf{X}=\{\mathbf{x}_M,\cdots,\mathbf{x}_2,\mathbf{x}_1\}$ is generated by $M$ successive steps in imaginary-time, from $\mathbf{x}_1$ to $\mathbf{x}_M$. Beginning with the partition function $Z=\text{Tr}(\hat{\mathcal{B}}\hat{\mathcal{B}}\cdots\hat{\mathcal{B}})$ where $\hat{\mathcal{B}}=e^{-\Delta\tau\hat{H}}$, we replace $\hat{\mathcal{B}}$ by Eq.~(\ref{eq:HSTransform}) one at a time from the right to the left. After $\ell$ such steps, the partial paths $\{\mathbf{x}_{\ell},\cdots,\mathbf{x}_{2},\mathbf{x}_{1}\}$ are constructed and the partition function can be written as $Z\simeq\sum_{\{\mathbf{x}_{\ell},\cdots,\mathbf{x}_{2},\mathbf{x}_{1}\}}\mathcal{P}_{\ell}(\{\mathbf{x}_{\ell},\cdots,\mathbf{x}_{2},\mathbf{x}_{1}\}, \hat{\mathcal{B}})$ with $\mathcal{P}_{\ell}$ as
\begin{equation}
\label{eq:PartConstrctPath}
\begin{split}
\mathcal{P}_{\ell}(\{\mathbf{x}_{\ell},&\cdots,\mathbf{x}_{2},\mathbf{x}_{1}\}, \hat{\mathcal{B}})  \\ &=\prod_{n=1}^{\ell}p(\mathbf{x}_n)\cdot\text{Tr}(\underbrace{\hat{\mathcal{B}}\cdots\hat{\mathcal{B}}
\hat{\mathcal{B}}}_{M-\ell}\underbrace{\hat{B}_{\ell}\cdots\hat{B}_2\hat{B}_1}_{\ell}).
\end{split}
\end{equation}
If we include all possible values of $\mathbf{x}_{\ell}$, all complete paths will be generated after $M$ steps, and we recover the full path-integral representation of the partition function in Eq.~(\ref{eq:PartitionFunc1}).

Let us consider a case with a particular partial path $\{\mathbf{x}_{\ell},\cdots,\mathbf{x}_{2},\mathbf{x}_{1}\}$ of length $\ell$, when $\mathcal{P}_{\ell}(\{\mathbf{x}_{\ell},\cdots,\mathbf{x}_{2},\mathbf{x}_{1}\}, \hat{\mathcal{B}})=0$. Mathematically this is equivalent to
\begin{equation}
\label{eq:PathNode}
\sum_{\{\mathbf{x}_{M},\cdots,\mathbf{x}_{\ell+1}\}}\prod_{n=1}^{M}p(\mathbf{x}_n)\cdot\text{Tr}(\hat{B}_M\cdots
\hat{B}_{\ell+1}\hat{B}_{\ell}\cdots\hat{B}_2\hat{B}_1)=0\,,
\end{equation}
where we have replaced the remaining $(M-\ell)$ many-body operators  $\hat{\mathcal{B}}$ in Eq.~(\ref{eq:PartConstrctPath}) by the corresponding $\hat{B}(\mathbf{x})$ operators with auxiliary fields, using Eq.~(\ref{eq:HSTransform}). Equation~(\ref{eq:PathNode}) implies that the contributions of all complete paths whose first $\ell$ elements are the particular partial path $\{\mathbf{x}_{\ell},\cdots,\mathbf{x}_{2},\mathbf{x}_{1}\}$ simply cancel in the partition function $Z$, as the summation over all possible $\{\mathbf{x}_{M},\cdots,\mathbf{x}_{\ell+2},\mathbf{x}_{\ell+1}\}$ is equal to zero.

This observation allows us to divide all the complete paths (or the auxiliary-field space) into two categories  according to their contributions to the partition function $Z$: an antisymmetric part and the residual part. All the complete paths containing the partial path $\{\mathbf{x}_{\ell},\cdots,\mathbf{x}_{2},\mathbf{x}_{1}\}$ discussed above belong to the antisymmetric category. Complete paths in this category
share the common characteristic that $\mathcal{P}_{\ell}$ defined in Eq.~(\ref{eq:PartConstrctPath}) encounters $\mathcal{P}_{\ell}=0$ for at least one $\ell\in[1,M]$. Statistically, the complete paths belonging to the antisymmetric category are "noise", since the contributions of such paths cancel in $Z$. These "noise" paths increasingly dominate the auxiliary-field space with lowering temperature; they contribute to the exponential growth in the variance of the numerical results. On the other hand, complete paths belonging to the other category, the residual part, are responsible for the actual signal in the
partition function; their contributions vanish exponentially with the length of the path. The overall effect gives rise to the behavior of the sign average discussed in Sec.~\ref{sec:SignProblem}, namely an exponential decay of signal-to-noise ratio that is the signature of the sign problem.

The recognition of the different categories of paths provides an understanding of the origin of the sign problem. It shows that a complete path contributes if and only if all of the following $M$ conditions are satisfied in the generation of the path:
\begin{equation}
\label{eq:CPMCContrExact}
\mathcal{P}_\ell(\{\mathbf{x}_{\ell},\cdots,\mathbf{x}_{2},\mathbf{x}_{1}\},\hat{\mathcal{B}}) >0, \hspace{0.7cm} \ell=1,2,\cdots, M.
\end{equation}
The constraints in Eq.~(\ref{eq:CPMCContrExact}) represent an absorbing boundary condition (BC) at $\mathcal{P}_{\ell}=0$ with increasing $\ell$. Under this BC, the probability distribution of the generated paths vanishes smoothly close to the boundary. Thus application of these constraints will eliminate all noise paths while keeping the contributing ones. Since the absorbing BC filters out the antisymmetric part in auxiliary-field space, the algorithm remains exact. This is the basic idea of the FT-AFQMC method.

To realize the FT-AFQMC method numerically, several additional issues must be addressed. First, a finite $\Delta\tau$ is always used in practical simulations, which makes it difficult to determine $\mathcal{P}_{\ell}=0$ precisely. Since $\mathcal{P}$ is continuous in $\ell$ only at the $\Delta\tau\to0$ limit, the
random walk will typically cross the absorbing boundary ``in between'' two successive time steps. This problem turns out to be straightforward to handle. In the simplest approximation, the absorbing boundary can be identified by ${\ell}$ corresponding to the first passage to $\mathcal{P}_{\ell}<0$. Alternatively, a higher order approach called mirror correction~\cite{Shiwei1995,Shiwei1997,Shiwei2000} can be applied, which introduces a finite probability to terminate the path at $(\ell-1)$ for the case when $\mathcal{P}_{\ell-1}>0$ and $\mathcal{P}_{\ell}<0$. These two approaches locate the absorbing boundary with controllable systematic errors of $\mathcal{O}(\Delta\tau)$ and $\mathcal{O}[(\Delta\tau)^2]$, respectively, which vanish as $\Delta\tau\to0$ and can be extrapolated away with finite $\Delta\tau$ calculations.

A second, much more significant issue is that $\hat{\mathcal{B}}$ in Eq.~(\ref{eq:CPMCContrExact}) is a many-body operator, which makes the exact calculation of $\mathcal{P}_\ell(\{\mathbf{x}_{\ell},\cdots,\mathbf{x}_{2},\mathbf{x}_{1}\},\hat{\mathcal{B}})$ in Eq.~(\ref{eq:PartConstrctPath}) a many-body problem. In practice, we replace it by a known trial propagator $\hat{B}_T=e^{-\Delta\tau\hat{H}_T}$, where $\hat{H}_T$ can be thought of as a trial Hamiltonian. This substitution causes a systematic error, since now the absorbing BC is only determined approximately. This is the constrained path approximation and it results in the only systematic error of the FT-AFQMC method. (All other numerical errors can be systematically removed, such as the finite-$\Delta\tau$ error, as discussed above, or population control error.) The constrained path approximation becomes exact if $\hat{B}_T$ is exact. In practice the trial propagator $\hat{B}_T$ is typically of single-particle form. In other words the trial $\hat{H}_T$ is often a single-particle Hamiltonian: $\hat{H}_T=\sum_{ij,\sigma\sigma^{\prime}}(\mathbf{H}_T)_{i\sigma,j\sigma^{\prime}} c_{i\sigma}^+c_{j\sigma^{\prime}}$, where $\mathbf{H}_T=\{(\mathbf{H}_T)_{i\sigma,j\sigma^{\prime}}\}$ is a $2N_s\times 2N_s$ Hermitian matrix. Using the trial $\hat{B}_T$, the $\mathcal{P}_{\ell}$ in Eq.~(\ref{eq:PartConstrctPath}) can be rewritten in the operator and matrix determinant forms as
\begin{equation}
\label{eq:PartConstrctPathTrialBT}
\begin{split}
\mathcal{P}_{\ell}^T&(\{\mathbf{x}_{\ell},\cdots,\mathbf{x}_{2},\mathbf{x}_{1}\})
= \mathcal{P}_{\ell}(\{\mathbf{x}_{\ell},\cdots,\mathbf{x}_{2},\mathbf{x}_{1}\},\hat{B}_T)
 \\ &=\prod_{n=1}^{\ell}p(\mathbf{x}_n)\cdot\text{Tr}(\underbrace{\hat{B}_T\cdots\hat{B}_T
\hat{B}_T}_{M-\ell}\hat{B}_{\ell}\cdots\hat{B}_2\hat{B}_1) \\
&=\prod_{n=1}^{\ell}p(\mathbf{x}_n)\cdot\text{det}\Big[\mathbf{I}_{2N_s}+\Big(\prod_{n=1}^{M-\ell}\mathbf{B}_T\Big)
\mathbf{B}_{\ell}\cdots\mathbf{B}_2\mathbf{B}_1\Big], \\
\end{split}
\end{equation}
where the matrix $\mathbf{B}_T=e^{-\Delta\tau \mathbf{H}_T}$. Then the constraints in Eq.~(\ref{eq:CPMCContrExact}) are replaced by the following
\begin{equation}
\label{eq:CPMCContrTrialBT}
\mathcal{P}_\ell^T(\{\mathbf{x}_{\ell},\cdots,\mathbf{x}_{2},\mathbf{x}_{1}\}) >0, \hspace{0.7cm} \ell=1,2,\cdots, M.
\end{equation}
which is imposed successively in imaginary time from $\ell=1$ to $M$.

\subsection{Random walk with importance sampling}
\label{sec:ImportSample}

We next discuss an algorithm for sampling the paths efficiently while imposing the constraints in Eq.~(\ref{eq:CPMCContrTrialBT}). Our goal is to generate complete paths (configurations) $\mathbf{X}$ which both satisfy the constraints and are distributed according to $D(\mathbf{X})$. We construct such contributing paths following exactly the process outlined in the thought experiment in Sec.~\ref{sec:FTCPMCApprox}. Instead of sweeping through the entire path for updates as in DQMC method, we construct the complete path $\mathbf{X}=\{\mathbf{x}_M,\cdots,\mathbf{x}_2,\mathbf{x}_1\}$ from $\mathbf{x}_1$ to $\mathbf{x}_M$ using a branching random walk, while imposing the constraints in Eq.~(\ref{eq:CPMCContrTrialBT}).

To introduce importance sampling into the random walk process to improve sampling efficiency, we note that the partition function can be rewritten as
\begin{equation}
\label{eq:CPMCPart000}
Z \simeq \sum_{\{\mathbf{x}_M,\cdots,\mathbf{x}_2,\mathbf{x}_1\}}\frac{\mathcal{P}_M^T}{\mathcal{P}_{M-1}^T}\frac{\mathcal{P}_{M-1}^T}{\mathcal{P}_{M-2}^T}\cdots
\frac{\mathcal{P}_2^T}{\mathcal{P}_1^T}\frac{\mathcal{P}_1^T}{\mathcal{P}_0^T} \mathcal{P}_0^T,
\end{equation}
where $\mathcal{P}_\ell^T$ is a short-hand for the full expression in Eq.~(\ref{eq:PartConstrctPathTrialBT}), and $\mathcal{P}_M^T=D(\mathbf{X})$. Starting from $\mathcal{P}_0^T=\text{det}(\mathbf{I}_{2N_s}+\prod_{n=1}^M\mathbf{B}_T)>0$, we first use $\lambda(\mathbf{x}_1)=\max[\mathcal{P}_1^T/\mathcal{P}_0^T,0]$ to construct a normalized probability density function $\eta(\mathbf{x}_1)=\lambda(\mathbf{x}_1)/\sum_{\mathbf{x}_1^\prime}\lambda(\mathbf{x}_1^\prime)$. We then draw a sample for $\mathbf{x}_1$ from $\eta(\mathbf{x}_1)$, and assign the normalization factor $\sum_{\mathbf{x}_1^\prime}\lambda(\mathbf{x}_1^\prime)$ as a weight of the newly sampled path. Note that the constraints in Eq.~(\ref{eq:CPMCContrTrialBT}) for $\ell=1$ have automatically been implemented by our choice of the probability density function. We then repeat the same procedure from $\mathbf{x}_2$ to $\mathbf{x}_M$. At the ${\ell}$-th step, with the partial path $\{\mathbf{x}_{\ell-1},\cdots,\mathbf{x}_{2},\mathbf{x}_1\}$ already constructed, we use the conditional PDF $\eta(\mathbf{x}_\ell)=\lambda(\mathbf{x}_{\ell})/\sum_{\mathbf{x}_{\ell}^\prime}\lambda(\mathbf{x}_{\ell}^\prime)$ to sample $\mathbf{x}_{\ell}$, where $\lambda(\mathbf{x}_\ell)=\max[\mathcal{P}_{\ell}^T/\mathcal{P}_{\ell-1}^T,0]$. The weight of the path is multiplied by the normalization $\sum_{\mathbf{x}_{\ell}^\prime}\lambda(\mathbf{x}_{\ell}^\prime)$.

In the calculation, we carry an ensemble of $N_{\mathbf{X}}$ samples and propagate them in parallel. Each sample is called a random walker, and the random walk is  carried out for $M$ steps. During the random walk, the walker weights can fluctuate and a population control procedure is applied periodically, as discussed in the next section. At the end of the random walk, $N_{\mathbf{X}}$ complete paths are obtained, given by $\mathbf{X}_k$ with weight $w_k$ for $k=1$ to $N_{\mathbf{X}}$. They provide Monte Carlo samples of a modified probability density function $W^{\rm c}(\mathbf{X})$ as defined in Eq.~(\ref{eq:W_X}) but with $D(\mathbf{X})$ replaced by $D^{\rm c}(\mathbf{X})$, where the superscript `c' means `under the constraint' of Eq.~(\ref{eq:CPMCContrTrialBT}). Observables can then be computed as in Eq.~(\ref{eq:PhyObs}), which is further discussed in the next section. We repeat the $M$-step random walk  procedure as needed to reach the desired statistical accuracy.

The FT-AFQMC method, as is now evident, does not use the Markov-Chain Monte Carlo employed in DQMC. The choice is driven by the difficulty in imposing the constraints in the path-integral formalism in DQMC. The constraining conditions are non-local in imaginary-time, since the condition at the $\ell$-th step depends on the path history from $1$ to $(\ell-1)$, or alternatively from $M$ to $(\ell+1)$ depending on how one views the reference point (or sweeping direction in the path sampling). This means that the sampling could get ``stuck'' with a configuration which violates the absorbing boundary condition. As we see from the analysis in Sec.~\ref{sec:FTCPMCApprox}, this will occur with higher and higher probability as the path becomes longer (lower temperature). The one-directional random walk with branching adopted in FT-AFQMC, which is similar to ZT-AFQMC, solves this problem.

In the discussion so far, we have implied that trial Hamiltonian $\hat{H}_T$ (or trial propagator $\hat{B}_T$) has no imaginary-time dependence. (Note that the time-dependence of $\hat{B}_T$ should be distinguished from that of the constraint; because of the product form in Eq.~(\ref{eq:PartConstrctPathTrialBT}), the constraint is time-dependent even if $\hat{B}_T$ is not.) It is straightforward to generalize the procedure to an imaginary-time dependent $\hat{H}_T$.

\subsection{Implementation issues for numerics}
\label{sec:NumericsImplement}

In this section we discuss several issues in a general implementation of the FT-AFQMC method for correlated fermion systems. They include numerical stablization, growth estimator and population control, measurements, and the form of $\hat{H}_T$ and its implementation. Additional details are provided in Appendix~\ref{sec:FTCPMCOtherDetail}

Finite-temperature AFQMC calculations, like DQMC, are more challenging to keep numerically stable than zero-temperature calculations. The instability grows more severe at lower temperatures, and is caused by numerical round-off errors~\cite{White1989,LOH2005,Tomas2012} from the multiplication of many $\mathbf{B}$ matrices. In our calculations, we use the column-pivoted QR algorithm to stabilize the matrix products and we also implement a very stable way~\cite{LOH2005,Tomas2012} to calculate the single-particle Green's function matrix shown in Appendix~\ref{sec:FTCPMCOtherDetail}. Combining them, we can access temperatures as low as $1/80$ in units of the hopping parameter (or $1/20$ in units of Fermi energy).

In the FT-AFQMC method, the weights of walkers, $w_k$, can fluctuate. The procedure to keep these fluctuations under control and maintain statistical accuracy is similar to that used in ground-state AFQMC calculations~\cite{Shiwei1997} and in diffusion Monte Carlo calculations~\cite{Hammond1994}. Several schemes are possible whose details differ somewhat and can affect the statistical accuracy and population control bias, but the effects are negligible (much smaller than the statistical errors) in our calculations with ${\mathcal{O}}(10^3$-$10^4)$ random walkers. Here we will not make distinctions of the technical details
of the population control schemes, but simply describe the simplest approaches.

In our calculations, a fixed population of $N_{\mathbf{X}}$ random walkers are kept. The weights of the walkers are carried. We often use a combing algorithm, which periodically resets all weights to unity by resampling them according to the normalized distribution $w_k /\sum_kw_k$. Alternatively, we monitor the weights for large and small values which are pre-defined thresholds (e.g., $5.0$ and $0.2$, respectively). If the overall weight $\sum_k w_k$ systematically increases or decreases with imaginary time, we can adjust all weights by a constant factor as needed, similar to the growth estimator of ground-state calculations~\cite{Shiwei1997}. Walkers with large weights are duplicated and those with small weights are eliminated with the appropriate probability (to maintain a statistically identical population).

The measurement of observables is quite straightforward after the contributing paths are generated with importance sampling. In the branching random walker formulation, every single random walker generates a contributing path $\mathbf{X}_k$ with the final overall weight $w_k$ as its contribution to the partition function $Z$. Thus the physical observable $\langle\hat{O}\rangle$ can be computed as
\begin{equation}
\label{eq:CPMCObsHere}
\langle\hat{O}\rangle = \frac{\sum_{k=1}^{N_{\mathbf{X}}}\omega_k O_k}{\sum_{k=1}^{N_{\mathbf{X}}}\omega_k},
\end{equation}
where $O_k$ is the measurement result within the configuration $\mathbf{X}_k$. Since the contributing path is generated from $\ell=1$ to $\ell=M$ time slices, the simplest way to calculate $O_k$ is to measure it at the $M$-th time slice (or $\tau=\beta$):
\begin{equation}
\label{eq:OkExpress}
O_k = \frac{\text{Tr}(\hat{O}\hat{B}_M\hat{B}_{M-1}\cdots\hat{B}_2\hat{B}_1)}
{\text{Tr}(\hat{B}_M\hat{B}_{M-1}\cdots\hat{B}_2\hat{B}_1)}\,.
\end{equation}
The evaluation of $O_k$ in Eq.~(\ref{eq:OkExpress}) is the same as discussed in  Sec.~\ref{sec:DQMCBasics} for DQMC. Although imaginary-time translational invariance is broken in FT-AFQMC, it is reasonable to expect that measurements at different times are of comparable quality and will become more equivalent with better trial $\hat{B}_T$. We generally take the average of the multiple measurements at different imaginary times, which can be easily achieved by wrapping the $\hat{B}_{\ell}$'s after the complete path has been sampled. The extra computational cost is often more than compensated for by the the gain in statistical accuracy. Furthermore, we find that the time-averaged results, which partially recovers translational invariance in imaginary-time, tend to have smaller systematic error than the single-time measurement at $\beta$. The procedure for measuring imaginary-time correlation functions is straightforward~\cite{AssaadEvertz2008,Ettore2016}.

We next comment on the operations involved in substituting the trial $\hat{B}_T$ by the interacting propagators in the sampling process. The Hamiltonian can be written as $\hat{H}=\hat{H}_0+\hat{H}_I=\hat{H}_T+(\hat{H}_I+\hat{H}_0-\hat{H}_T)$. The trial propagator is $\hat{B}_T=e^{-\Delta\tau\hat{H}_T}$, and we need to insert the $e^{-\Delta\tau(\hat{H}_I+\hat{H}_0-\hat{H}_T)}$ operator. Operationally, this means replacing $\hat{B}_T$ by $\hat{B}=\hat{B}_I(\mathbf{x})\hat{B}_K$ at every time slice. The overall weight of the generated path comes from two parts, the importance sampling of the auxiliary fields and the ratio of the determinants when changing $\mathbf{B}_T$ to $\mathbf{B}_K$. In the special case when we choose $\hat{H}_T=\hat{H}_0$, as is the case when a restricted Hartree-Fock form is used, there is an additional simplification, and we only need to insert the corresponding $\mathbf{B}_I(\mathbf{x}_{\ell})$ matrix. Further details and the formulas for $\mathcal{P}_{\ell}^T/\mathcal{P}_{\ell-1}^T$ are provided in Appendix~\ref{sec:FTCPMCOtherDetail}.

\section{FT-AFQMC with self-consistent constraint}
\label{sec:FTCPMCTralSelf}

In this section, we present a self-consistent constraint in FT-AFQMC, after first carrying out a systematic study of the effect of different choices of mean-field trial density matrices. For concreteness, we will use the doped two-dimensional repulsive Hubbard model as an example. However, much of the discussion can be generalized to other Hamiltonians, including realistic electronic Hamiltonians under the phaseless formalism of the constraint~\cite{Shiwei2003}.

\subsection{The trial density matrix: illustration in the Hubbard model}
\label{sec:RHFUHFTrial}

The one-band Hubbard model is a representative model for studying correlation effects of interacting electrons. The model Hamiltonian $\hat{H}=\hat{H}_0+\hat{H}_I$ is as follows
\begin{equation}
\label{eq:HubbardModel}
\begin{split}
\hat{H} = &-t\sum_{\langle ij\rangle\sigma}(c_{i\sigma}^+c_{j\sigma}+c_{j\sigma}^+c_{i\sigma}) + \mu\sum_{i} (\hat{n}_{i\uparrow} + \hat{n}_{i\downarrow} )  \\
 &+ U\sum_{i}\Big( \hat{n}_{i\uparrow}\hat{n}_{i\downarrow} - \frac{\hat{n}_{i\uparrow} + \hat{n}_{i\downarrow}}{2} \Big),
\end{split}
\end{equation}
where $\hat{n}_{i\sigma}=c_{i\sigma}^+c_{i\sigma}$ is the density operator on the lattice site $i=(i_x,i_y)$. The nearest-neighbor hopping $t$, on-site Coulomb interaction $U$ and chemical potential $\mu$ are model parameters. In this work, we focus on repulsive interaction, $U>0$. The Hamiltonian above is written such that $\mu=0$ gives half-filling, with $\mu>0$ for hole doping and $\mu<0$ for electron doping. The overall electron density is given as $n=(N_{\uparrow}+N_{\downarrow})/N_s$, and the hole density, or doping, is then $(1-n)$.

To choose a trial density matrix, the simplest $\hat{H}_T$ to consider is the restricted Hartree-Fock (RHF) type:
\begin{equation}
\label{eq:HubbardRHFTrial}
\hat{H}_T = -t\sum_{\langle ij\rangle\sigma}(c_{i\sigma}^+c_{j\sigma}+c_{j\sigma}^+c_{i\sigma}) + \sum_{i} \mu_{i,T}(\hat{n}_{i\uparrow} + \hat{n}_{i\downarrow} ),
\end{equation}
where $\mu_{i,T}$ are free parameters. In this work, we only consider the simplified case of $\mu_{i,T}=\mu_T$, namely keeping translational invariance. Taking $\mu_T=\mu$ simply gives $\hat{H}_T=\hat{H}_0$. More importantly, $\mu_T$ can be used to tune the electron filling $\langle\hat{n}\rangle_T$ for the trial Hamiltonian.

Another trial Hamiltonian for Hubbard model is the unrestricted Hartree-Fock (UHF) type:
\begin{equation}
\label{eq:HubbardUHFTrial}
\begin{split}
\hat{H}_T = -t\sum_{\langle ij\rangle\sigma}&(c_{i\sigma}^+c_{j\sigma}+c_{j\sigma}^+c_{i\sigma}) \\
&+ \sum_{i\sigma}\Big[ U_{T}\Big(\langle \hat{n}_{i\bar{\sigma}}\rangle-\frac{1}{2} \Big) +\mu_{T} \Big]\hat{n}_{i\sigma},
\end{split}
\end{equation}
where $\bar{\sigma}$ denotes the opposite of $\sigma$, and $\{\langle\hat{n}_{i\uparrow}\rangle,\langle\hat{n}_{i\downarrow}\rangle,i=1,2,\cdots,N_s\}$ are from a self-consistent solution with $U=U_T$. The $\mu_T$ parameter here is similar to that in the RHF trial Hamiltonian in Eq.~(\ref{eq:HubbardRHFTrial}). We have omitted a constant term which only affects the overall weights as discussed in the previous section.

Both the RHF and UHF trial Hamiltonians resemble (or are identical) to the $\hat{H}_0$ term in the many-body Hamiltonian in Eq.~(\ref{eq:HubbardModel}). Thus computational simplifications exist in the updating procedure as discussed in the previous section. Of course in principle any other form of single-particle $\hat{H_T}$ can be used as trial Hamiltonian.

\begin{figure*}[!tp]
\centering
\includegraphics[width=1.7\columnwidth]{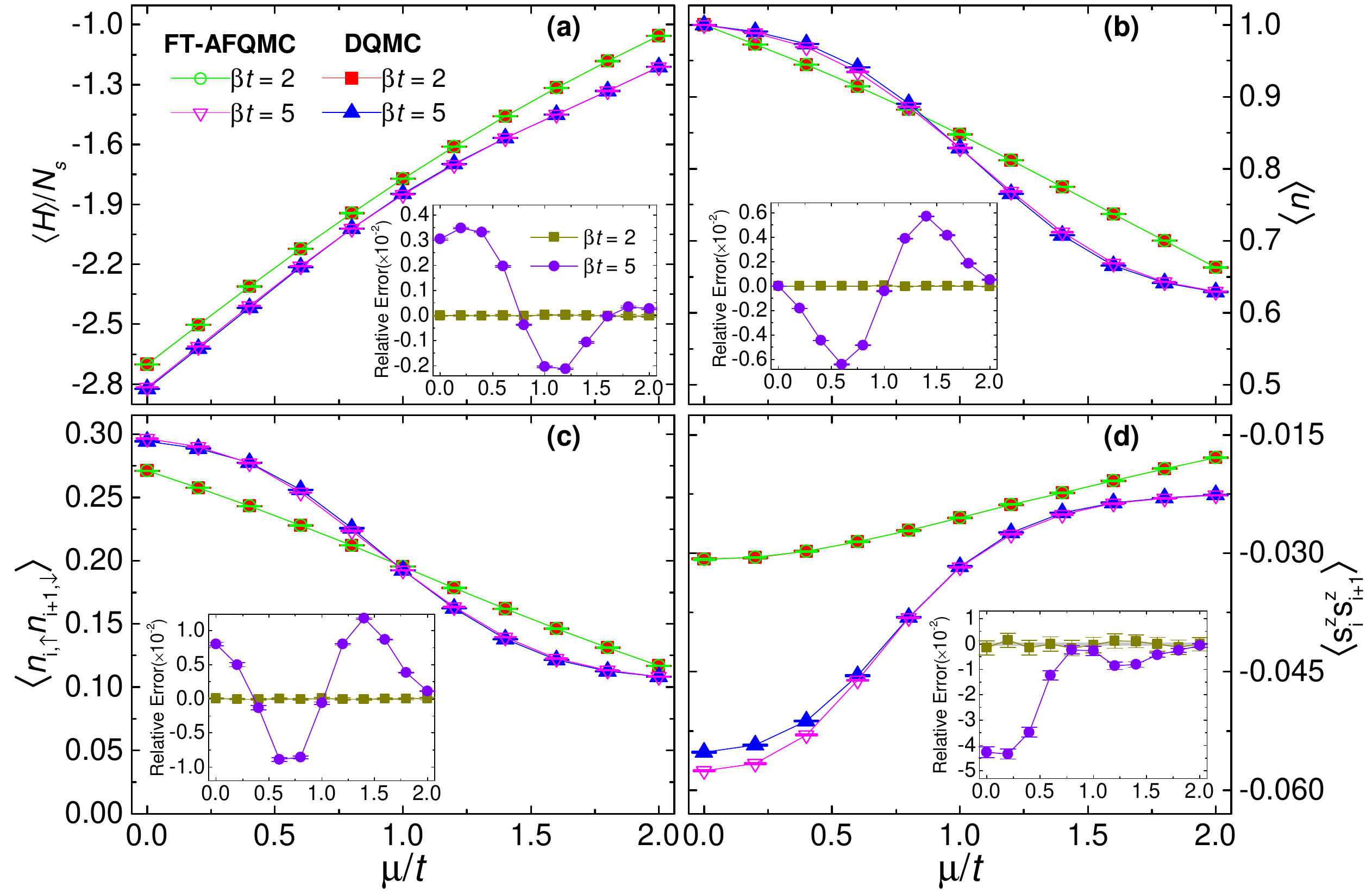}
\caption{\label{fig:RHFTrialVaryMu}(Color online) Benchmark of FT-AFQMC at high temperatures. Results are shown for Hubbard model under periodic boundary conditions, on a $4\times4$ system with $U/t=4$ and varying $\mu/t$, comparing FT-AFQMC with DQMC calculations at $\beta t=2$ and $\beta t=5$. The four panels give: (a) Total energy per site, (b) electron filling, (c) NN density-density correlation function $\langle\hat{n}_{i,\uparrow}\hat{n}_{i+1,\downarrow}\rangle$, and (d) NN spin-spin correlation function $\langle\hat{s}_i^z\hat{s}_{i+1}^z\rangle$. The relative errors of FT-AFQMC results are shown in the corresponding inset plots.}
\end{figure*}

\subsection{FT-AFQMC calculations with self-consistent procedure}
\label{sec:SCFUHFMethod}

The self-consistent procedure we develop here will allow us to improve the constraint based on feedback from the AFQMC calculation. It generalizes the method discussed in Ref.~\onlinecite{MingPu2016b} under the ZT-AFQMC framework to finite temperatures. We will optimize the trial Hamiltonian (or trial propagator) step by step iteratively through a sequence of FT-AFQMC calculation so as to reduce the error from the constraint.

For illustrating the method and testing our results, we will consider a more general form of the Hamiltonian of Eq.~(\ref{eq:HubbardModel}), by including an extra external potential term $\sum_{i\sigma}v_{i\sigma}\hat{n}_{i\sigma}$. This will allow us to add magnetic pinning fields~\cite{White2007,White2009,Assaad2013} to break translational invariance in illustrating the method and testing our results. With pinning fields, spin and charge correlations in periodic systems
can be probed by one-body spin and charge order parameters, whose measurements are straightforward from the single-particle Green's functions.

The basic idea of self-consistent procedure is to couple the QMC calculation to an independent-particle (IP) calculation. The IP calculation deals with the following Hamiltonian, including an external pinning field:
\begin{equation}
\label{eq:IPHamiltonian}
\begin{split}
\hat{H}_{\text{IP}} = &-t\sum_{\langle ij\rangle\sigma}(c_{i\sigma}^+c_{j\sigma}+c_{j\sigma}^+c_{i\sigma}) + \sum_{i\sigma}v_{i\sigma}\hat{n}_{i\sigma} \\
&+ \sum_{i\sigma} \Big[ U_{\text{eff}}\Big(\langle\hat{n}_{i\bar{\sigma}}\rangle-\frac{1}{2} \Big) +\mu_{\text{eff}} \Big]\hat{n}_{i\sigma},
\end{split}
\end{equation}
where $U_{\text{eff}}$ and $\mu_{\text{eff}}$ are tuning parameters, and the input densities $\{\langle\hat{n}_{i\uparrow}\rangle,\langle\hat{n}_{i\downarrow}\rangle\}$ are from a preceding QMC calculation. The solution is used to construct the input trial Hamiltonian for the next QMC calculation in the self-consistent procedure.

We outline the method in concrete steps:
\begin{enumerate}
\item
Start from a FT-AFQMC calculation using any typical choice of trial Hamiltonian $\hat{H}_T$, for example RHF or UHF.

\item
$\hat{H}_{\text{IP}}$ in Eq.~(\ref{eq:IPHamiltonian}) is solved using the densities obtained from the AFQMC calculation in the previous step: $\langle\hat{n}_{i\sigma}\rangle_{\text{QMC}}\to\langle\hat{n}_{i\sigma}\rangle_{\text{IP}}$. We vary $U_{\text{eff}}$ to find an optimal value, with which the computed densities {\it at the targeted temperature} are closest to the input from QMC, by minimizing the following function
\begin{equation}
\label{eq:ChiFunction}
\chi = \sqrt{\frac{1}{2N_s}\sum_{i\sigma}\Big( \langle\hat{n}_{i\sigma}\rangle_{\text{IP}} - \langle\hat{n}_{i\sigma}\rangle_{\text{\text{QMC}}} \Big)^2}.
\end{equation}

\item
The IP Hamiltonian $\hat{H}_{\text{IP}}$ with the optimal $U_{\text{eff}}$ and input densities from the previous AFQMC calculation is taken as trial Hamiltonian $\hat{H}_T$ to perform a new FT-AFQMC calculation.

\item
Return to step (2) or stop if convergence criteria is reached
\end{enumerate}

As mentioned, $\mu_{\text{eff}}$ in Eq.~(\ref{eq:IPHamiltonian}) is a tuning parameter similar to $\mu_T$ in the RHF and UHF trial Hamiltonians in Eq.~(\ref{eq:HubbardRHFTrial}) and Eq.~(\ref{eq:HubbardUHFTrial}). In our calculations, we tune $\mu_{\text{eff}}$ to make $\langle\hat{n}\rangle_{\text{IP}}$ equal to the desired electron filling of the many-body system.

Two somewhat different schemes can be used to obtain $T>0$ results at each electron filling $\langle\hat{n}\rangle$. We can perform the above self-consistent procedure at $T$ within FT-AFQMC calculations. Alternatively, we can first apply the self-consistent procedure using ZT-AFQMC method~\cite{MingPu2016b} to obtain the converged results of $U_{\text{eff}}$ and densities at $T=0$. Substituting these converged results into the IP Hamiltonian $\hat{H}_{\text{IP}}$ in Eq.~(\ref{eq:IPHamiltonian}), we can then take $\hat{H}_{\text{IP}}$ to generate a UHF trial Hamiltonian once for all to perform FT-AFQMC calculations. The temperature dependence of the UHF trial Hamiltonian only lies in $\mu_{\text{eff}}$ parameter, which is again tuned so that $\langle\hat{n}\rangle_{\text{IP}}$ match the filling $\langle\hat{n}\rangle$. We have done a careful comparison of these two schemes in the benchmark study. Both were found to yield accurate results with only slight differences visible at high and low temperatures. We will only present the results of finite-$T$ self-consistent calculations in the following.

\section{Benchmark results}
\label{sec:NumericalResults}

In this section we present benchmark results of FT-AFQMC calculations by comparing them with those from DQMC at $T>0$ and ED, DMRG, and  ZT-AFQMC at $T=0$. The results are divided into two parts. In the first part, we perform a detailed test of the accuracy of ``one-shot" FT-AFQMC calculations, using periodic supercells. Then in the second part we study the self-consistent procedure with FT-AFQMC, calculations are presented in the second part, with Hubbard model on a studying mostly systems with pinning fields. When applying the antiferromagnetic (AFM) pinning fields to an $L_x\times L_y$ supercells, we add a term $\sum_{i\sigma}v_{i\sigma}\hat{n}_{i\sigma}$ with $v_{i\uparrow}=-v_{i\downarrow}=(-1)^{i_y}h$ for both $i_x=1$ (and sometimes also $i_x=L_x$ as we will specify below) in Eq.~(\ref{eq:HubbardModel}). The supercell remains periodic along the $y$-direction and open along $x$-direction. With pinning fields, spin and charge correlations in periodic systems can be probed by one-body spin and charge order parameters, whose measurements are straightforward from the single-particle Green's functions.

We apply the symmetric decomposition $e^{-\Delta\tau\hat{H}}=e^{-\Delta\tau\hat{H}_0/2}e^{-\Delta\tau\hat{H}_U}e^{-\Delta\tau\hat{H}_0/2} + \mathcal{O}[(\Delta\tau)^3]$ and use the following discrete form~\cite{Hirsch1983} for Eq.~(\ref{eq:HSTransform})
\begin{equation}
\label{eq:HubbardHSTransf}
e^{-\Delta\tau U\big(n_{i\uparrow}n_{i\downarrow}-\frac{n_{i\uparrow}+n_{i\downarrow}}{2}\big)}=\sum_{x_i=\pm1}\frac{1}{2}e^{\lambda x_i(n_{i\uparrow}-n_{i\downarrow})}.
\end{equation}
in both DQMC and FT-AFQMC calculations. In Eq.~(\ref{eq:HubbardHSTransf}), $\lambda=\cosh^{-1}(e^{\Delta\tau U/2})$ for $U>0$ and $x_i$ is an auxiliary-field.

Conservative values of $\Delta\tau$, typically $0.05$ or $0.02$, are used, and the same $\Delta\tau$ is applied in DQMC and FT-AFQMC when comparing their numerical results. The population of random walkers ranges from $10^3$ to $10^4$. All the results we will present are from the multiple measurements along the contributing paths as discussed in Sec.~\ref{sec:NumericsImplement}, unless otherwise noted.

\subsection{Numerical results from FT-AFQMC calculations with specific trial Hamiltonians}
\label{sec:FTCPMCWithRHF}

We first benchmark the FT-AFQMC at high temperatures. The Hubbard model on a $4\times4$ square lattice with periodic boundary conditions will be used, for availability of exact results. We study $\beta t=2$ and $\beta t=5$, varying the $\mu$ parameter to examine different electron fillings. The sign problem is mild in these situations, and we obtain accurate results from DQMC calculations for comparison. In AFQMC, the RHF trial Hamiltonian of Eq.~(\ref{eq:HubbardRHFTrial}) is applied. We determine $\mu_T$ by the condition $\langle\hat{n}\rangle_T=\langle\hat{n}\rangle$, i.e., the electron filling of $\hat{H}_T$ is equal to the targeted filling of the many-body system, as defined by DQMC. The choice of  $\mu_T$ is further discussed below.

\begin{figure}[t]
\centering
\includegraphics[width=0.96\columnwidth]{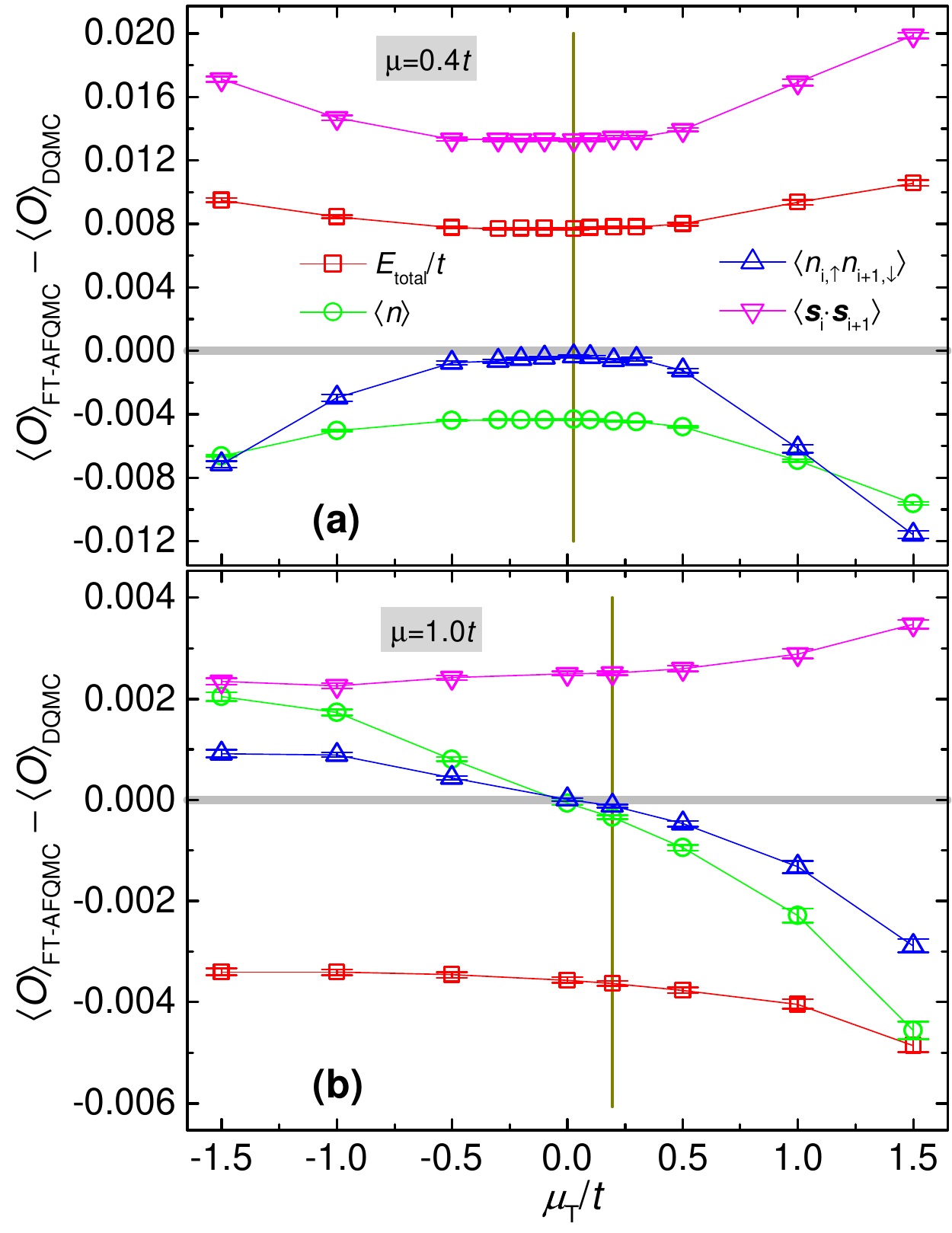}
\caption{\label{fig:RHFTrialMuTMu0410}(Color online) Systematic errors from FT-AFQMC calculations as the $\mu_T/t$ parameter in the applied RHF trial Hamiltonian $\hat{H}_T$ is varied. The system is the same as in Fig.~\ref{fig:RHFTrialVaryMu} at $\beta t=5$. (a) $\mu=0.4t$ and (b) $\mu=1.0t$ are studied as representative cases. Gray, horizontal line represents zero error and the dark yellow, vertical line stands for the $\mu_T/t$ value in Fig.~\ref{fig:RHFTrialVaryMu}, determined from $\langle\hat{n}\rangle_T=\langle\hat{n}\rangle_{\text{DQMC}}$.}
\end{figure}

Results for the energy, electron filling, density and spin correlations are shown in Fig.~\ref{fig:RHFTrialVaryMu} for both DQMC and FT-AFQMC. It is evident that the results from both methods are perfectly consistent at $\beta t=2$ for all $\mu$ parameters, with zero relative discrepancy within statistical error bars. This confirms that the FT-AFQMC results are exact at high temperature, even with the simplest RHF trial Hamiltonian. (For the $\beta t=2$ results shown, we actually took $\mu_T=\mu$, i.e., $\hat{H}_T=\hat{H}_0$. Tuning it with respect to the exact density yields indistinguishable results.) For $\beta t=5$, FT-AFQMC calculations with RHF trial Hamiltonian also generate quite accurate results, with the largest relative error $<1\%$ for total energy, electron filling, density correlation function and about $4\%$ for spin correlation function. For spin correlation function, the largest deviation appears at half-filling with $\mu=0$. This can be improved by using trial Hamiltonian with UHF and becomes essentially exact in ZT-AFQMC with generalized Hartree-Fock (GHF) trial state~\cite{Shihao2014,Chang2017}.

We next study more closely the role of $\mu_T$ in the RHF trial Hamiltonian, with additional FT-AFQMC calculations at $\beta t=5$. Two representative points are checked, $\mu=0.4t$ and $\mu=1.0t$, where relative errors shown in Fig.~\ref{fig:RHFTrialVaryMu} are close to the largest. We vary $\mu_T$ and study the corresponding systematic error of FT-AFQMC, $\langle\hat{O}\rangle_{\text{FT-AFQMC}}-\langle\hat{O}\rangle_{\text{DQMC}}$. For $\mu=0.4t$, as shown in Fig.~\ref{fig:RHFTrialMuTMu0410}(a), all the physical observables have the smallest systematic errors at the $\mu_T/t$ value determined from the $\langle\hat{n}\rangle_T=\langle\hat{n}\rangle_{\text{DQMC}}$ condition. That is not the case for $\mu=1.0t$ as shown in Fig.~\ref{fig:RHFTrialMuTMu0410}(b), for which the smallest errors for different observables are located at different $\mu_T/t$ values. Thus no  $\mu_T$ exists within the RHF trial Hamiltonian framework which ``optimizes'' the FT-AFQMC calculation in an absolute sense. However, at the position of $\mu_T/t$ determined from the exact electron filling condition, the systematic errors of all the quantities are very close to the minimum, validating our earlier choice. Furthermore, the errors all vary slowly (note the small scale of the errors in the plot) in a broad range of $\mu_T/t$ values, indicating that the constraint is not very sensitive to the details of the trial Hamiltonian.

\begin{figure}[t]
\centering
\includegraphics[width=0.96\columnwidth]{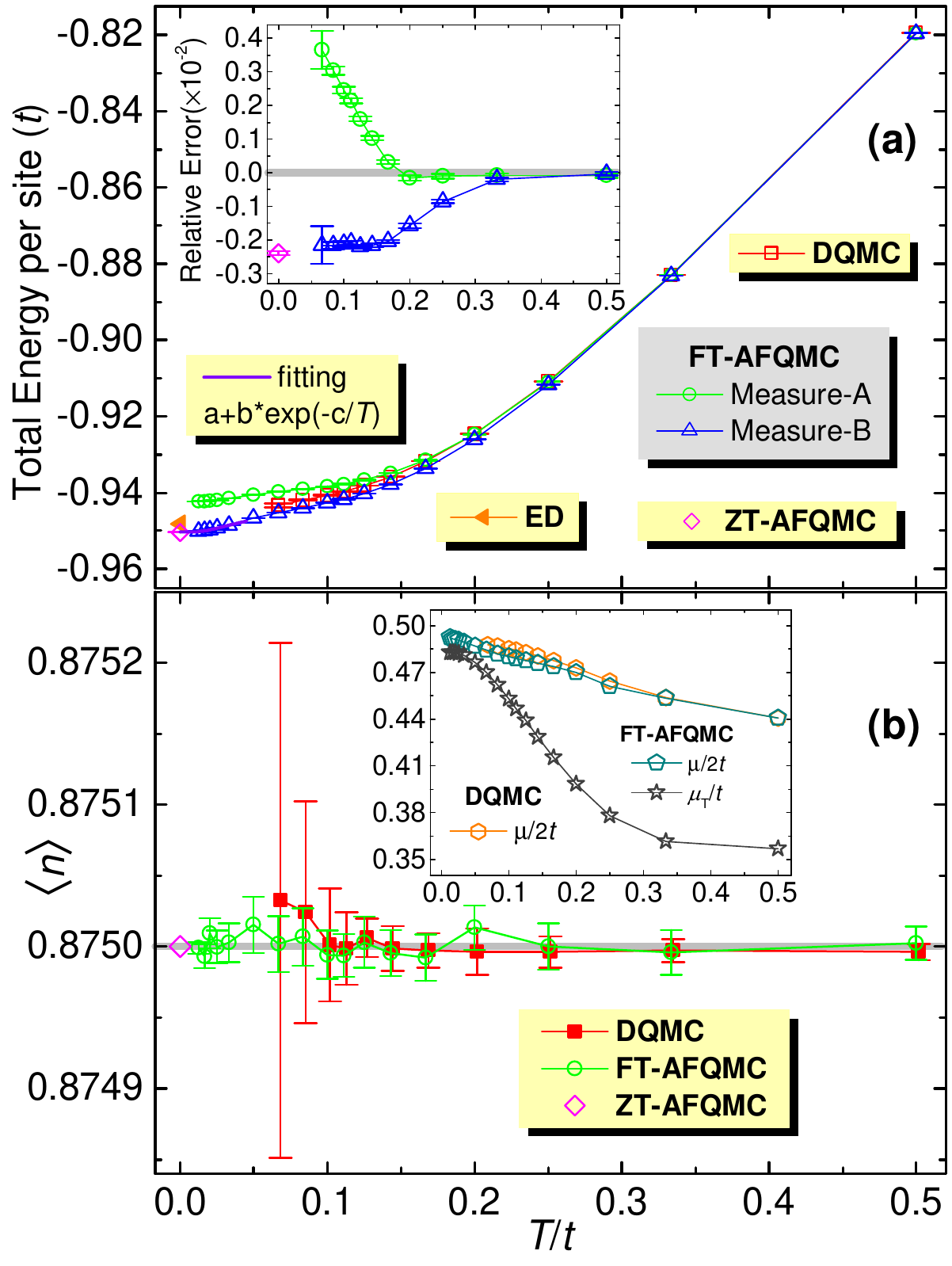}
\caption{\label{fig:RHF0875Pin}(Color online) Benchmark results over the complete range of temperatures. The system is a $4\times4$ square lattice with $U/t=4$ and AFM pinning fields $h=0.1t$ at $i_x=1$, targeting $\langle\hat{n}\rangle=0.875$. The computed total energy vs.~temperature is shown in (a), and the relative errors are shown with respect to DQMC for $T=t/15\sim t/2$ region and then with respect to ED at $T=0$\,K. Measure-A  denotes doing a single measurement at $\tau=\beta$ in FT-AFQMC, while Measure-B denotes averaging multiple measurements along the whole path. The computed density is shown in (b), with the numerically determined  $\mu$ parameter shown in the inset together with the $\mu_T$ parameter in the RHF trial Hamiltonian $\hat{H}_T$. }
\end{figure}

Now we study a broader temperature range and examine the behavior of the FT-AFQMC algorithm for accessing sufficiently low temperatures to approach the ground state. We focus on $U=4t$ with $1/8$ hole doping, i.e. $\langle\hat{n}\rangle=0.875$, and a temperature range from $T=t/2$ to $T=t/80$. Here we apply an AFM pinning field $h=0.1t$ on one edge, at $i_x=1$, with open boundary condition the along $x$ direction in the $4\times4$ system. All reference calculations, ED, ZT-AFQMC, and DQMC, are performed on the same Hamiltonian and parameter choices. The lowest temperature accessible here with DQMC is $T=t/15$, with average sign of $0.031$. We use the RHF trial Hamiltonian (with the same AFM pinning fields) in FT-AFQMC calculations.

The results of this benchmark are shown in Fig.~\ref{fig:RHF0875Pin}. In panel (a), energies computed from FT-AFQMC at high temperature, e.g, $\beta t=2$,
are indistinguishable from DQMC as seen earlier. As $T$ decreases, the systematic error becomes visible and tends to increase, with a maximum of about $0.2\%$ deviation at the lowest temperature available from DQMC, $T=t/15$. In the lower temperature region, we find that the energy with Measure-B obtained from FT-AFQMC fits accurately to an exponential, $E=E_0+be^{-c/T}$. The extrapolated value $E_0$ at $T=0$ is consistent with the result from ZT-AFQMC (measured with back propagation). This consistency unifies the FT- and ZT-AFQMC framework. Thus we expect a maximum relative error, as $T\rightarrow 0$, of $\sim 0.24\%$ as given by ZT-AFQMC, as shown in inset of Fig.~\ref{fig:RHF0875Pin}(a). This is very accurate, especially since it is obtained with the simplest RHF form of the trial Hamiltonian. The energy results from Measure-A (single measurement at $\tau=\beta$) and Measure-B (averaging multiple measurements along complete path) show significant differences at low temperatures, as seen in Fig.~\ref{fig:RHF0875Pin}(a) and the inset. The improvement from Measure-B at low temperatures is an effect of partially restored imaginary-time translational symmetry, as mentioned in Sec.~\ref{sec:NumericsImplement}. As seen in Fig.~\ref{fig:RHF0875Pin}(b), the numerically calculated overall filling $\langle\hat{n}\rangle$ from both DQMC and FT-AFQMC methods are indeed equal to the desired value $0.875$, within statistical error bars, across the entire range of temperature. The calculated electron fillings are the same for Measure-A and Measure-B, since the density operator commutes with the ${\hat B}$ used. As shown in inset of Fig.~\ref{fig:RHF0875Pin}(b), the $\mu$ parameters determined from DQMC and FT-AFQMC calculations, to produce the desired density, are in good agreement at high temperatures, but discrepancy is visible with decreasing $T$. This is a consequence of the constrained path approximation in FT-AFQMC. The $\mu_T$ parameter in the RHF trial Hamiltonian for AFQMC, determined via the density condition $\langle n\rangle_T=\langle n\rangle=0.875$, is also indicated in the plot.

\begin{figure*}
\centering
\includegraphics[width=0.670\columnwidth]{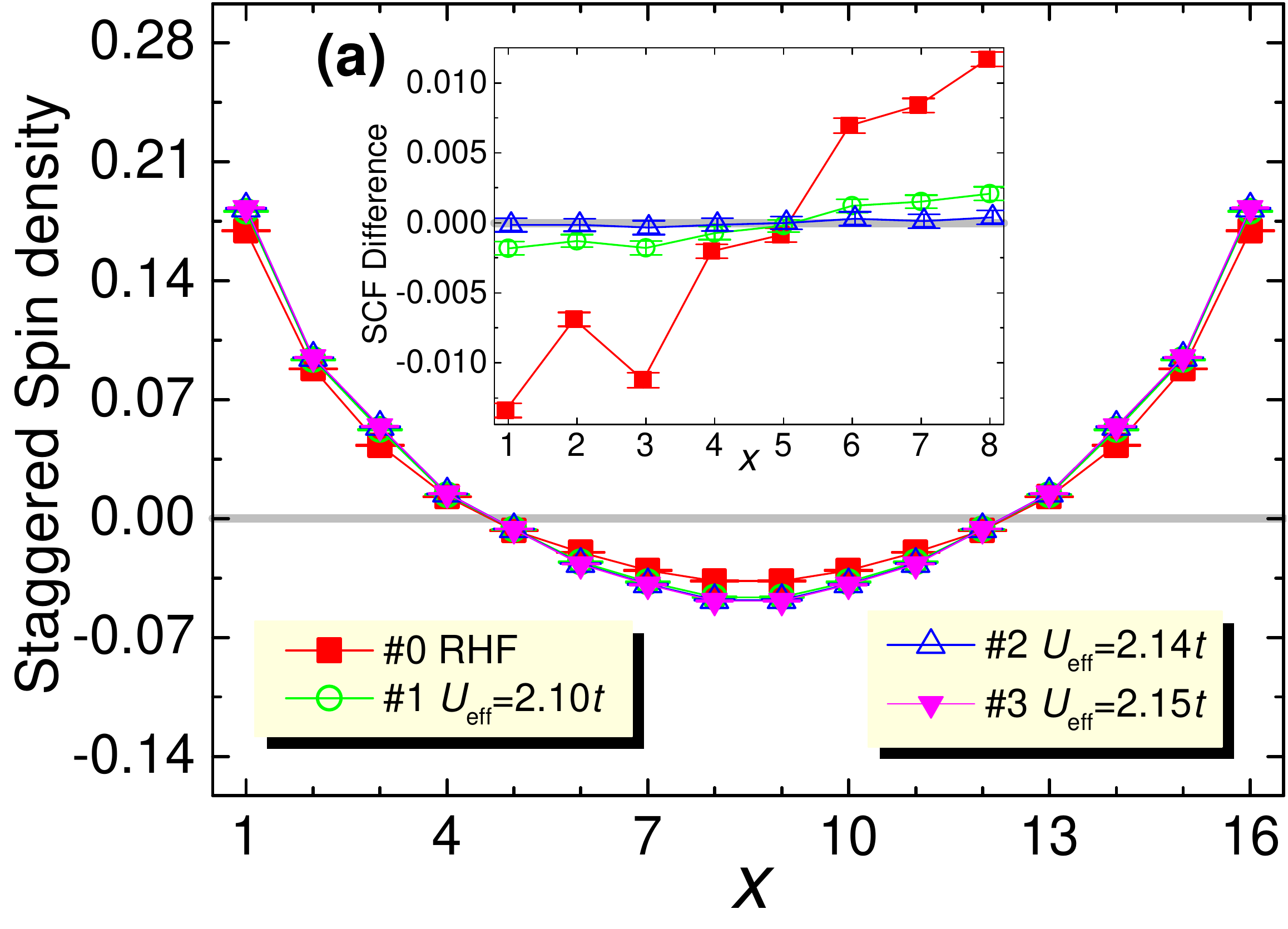}
\includegraphics[width=0.670\columnwidth]{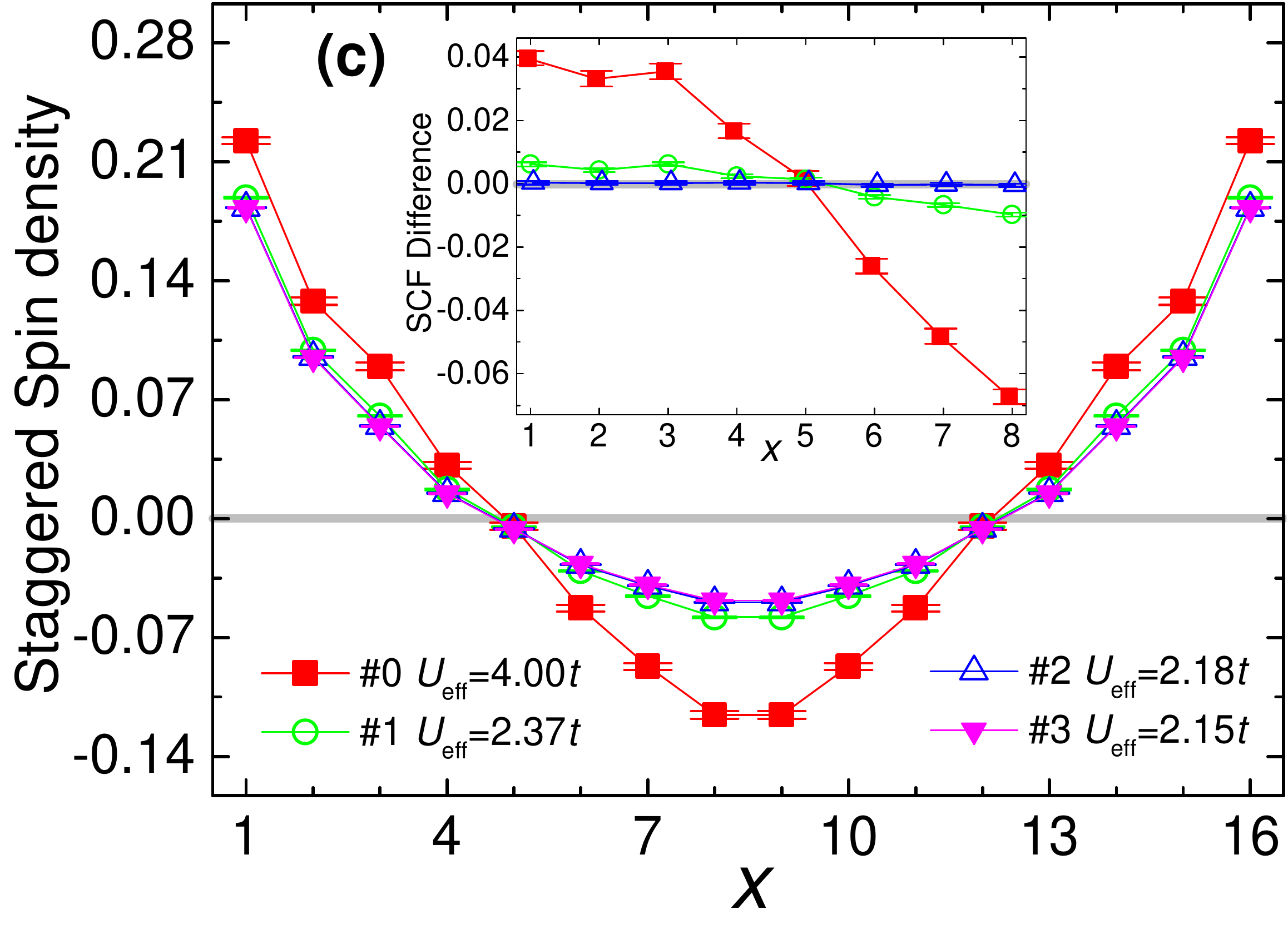}
\includegraphics[width=0.706\columnwidth]{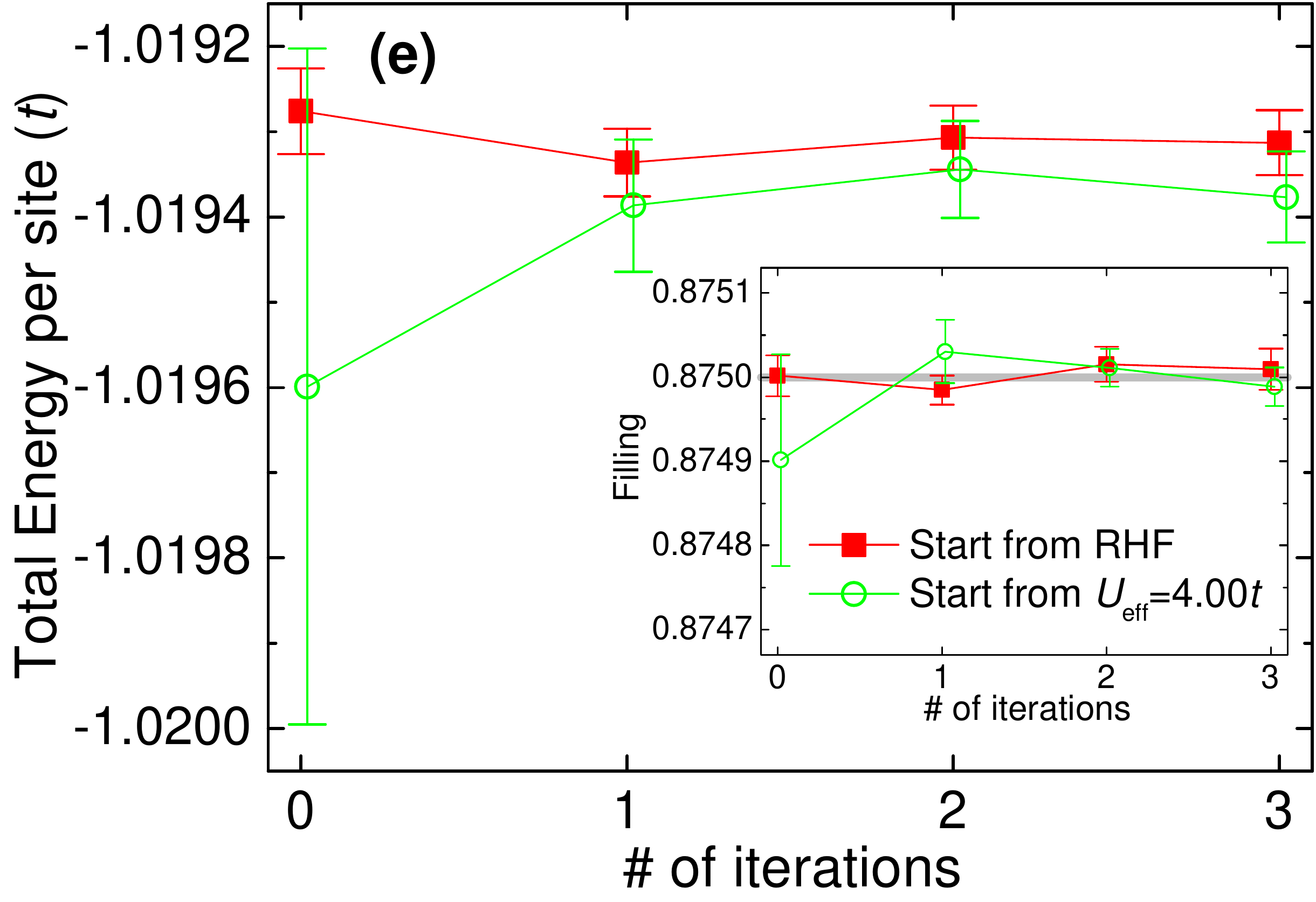} \\
\includegraphics[width=0.668\columnwidth]{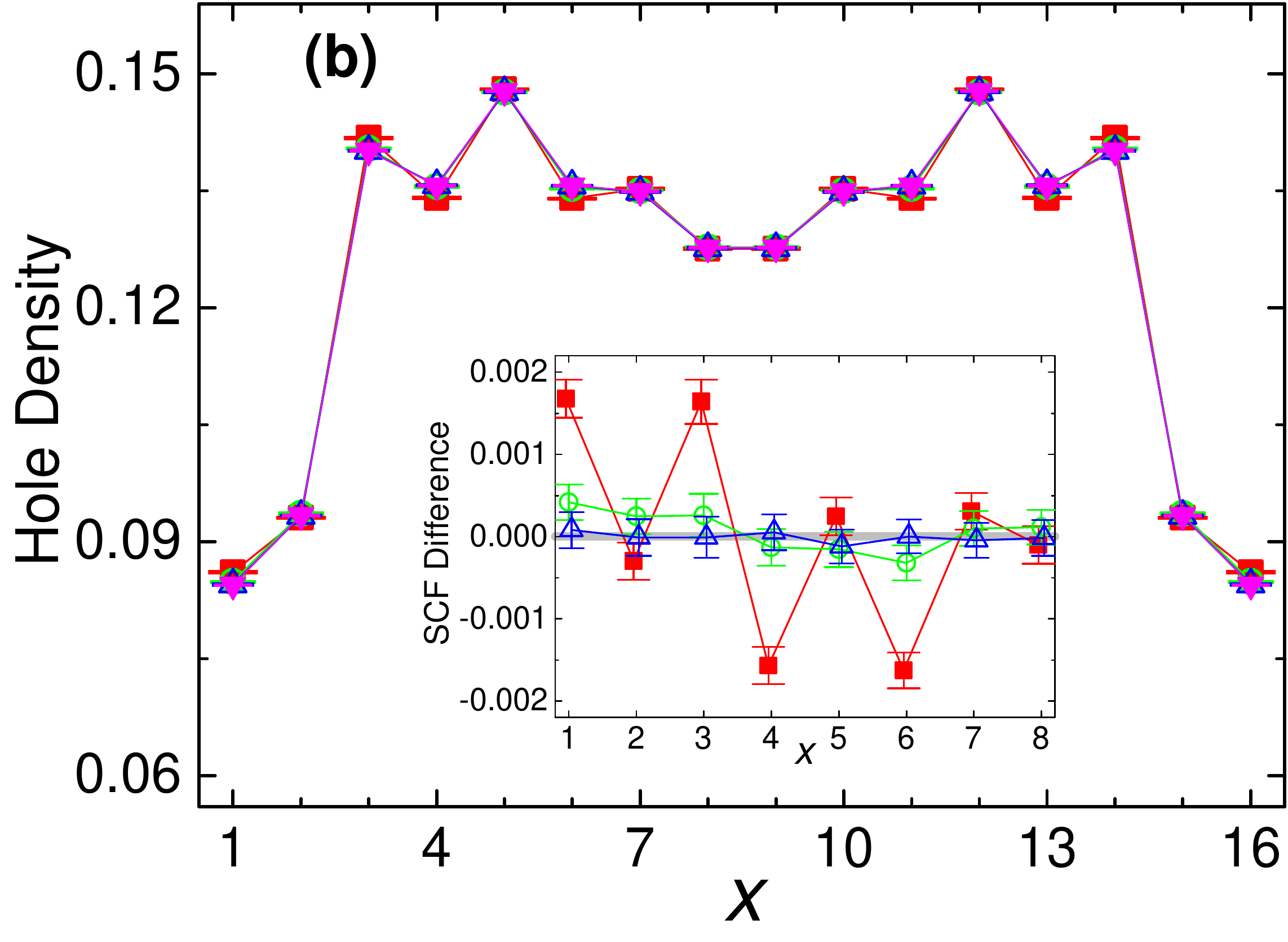}
\includegraphics[width=0.668\columnwidth]{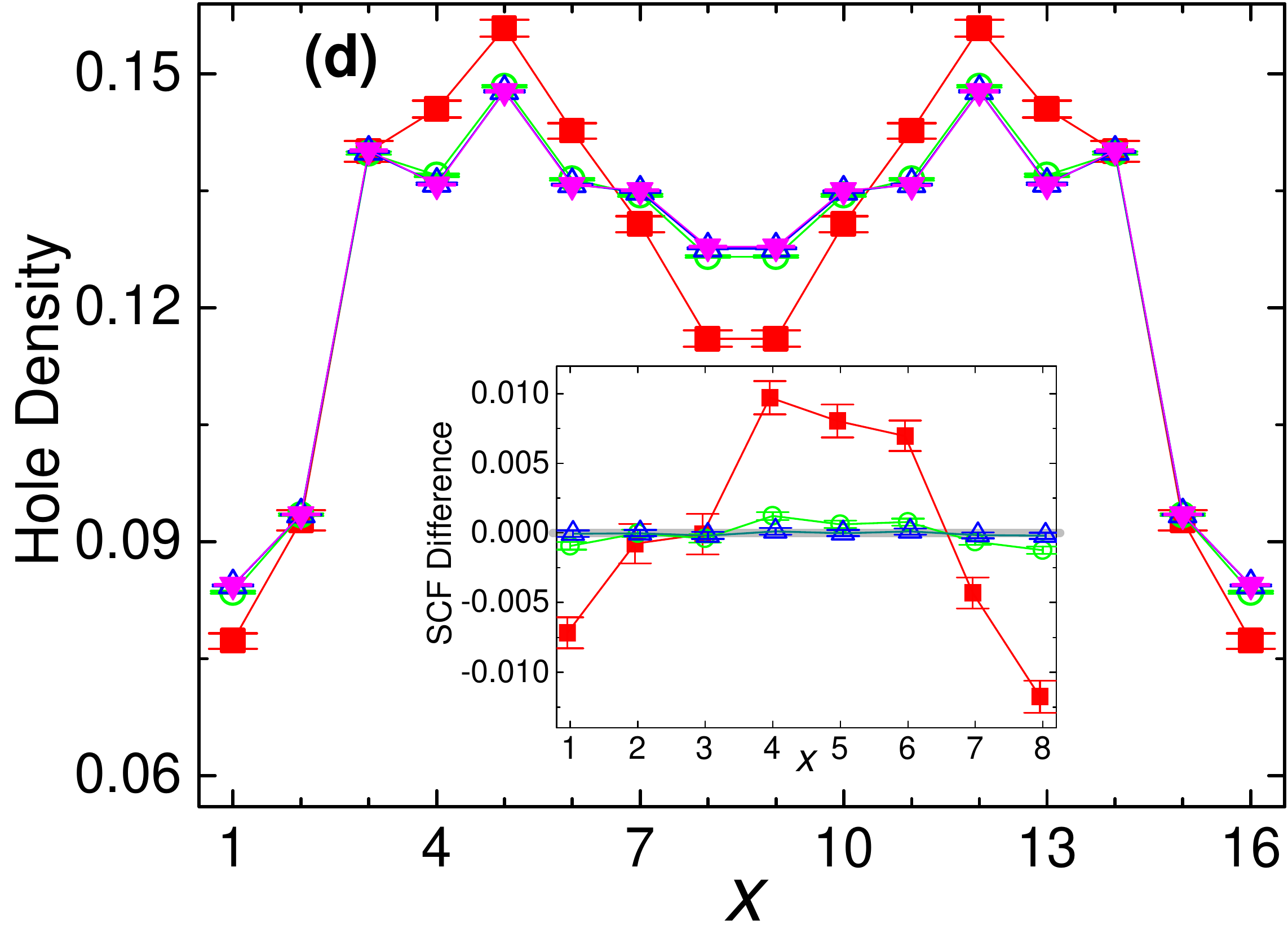}
\includegraphics[width=0.713\columnwidth]{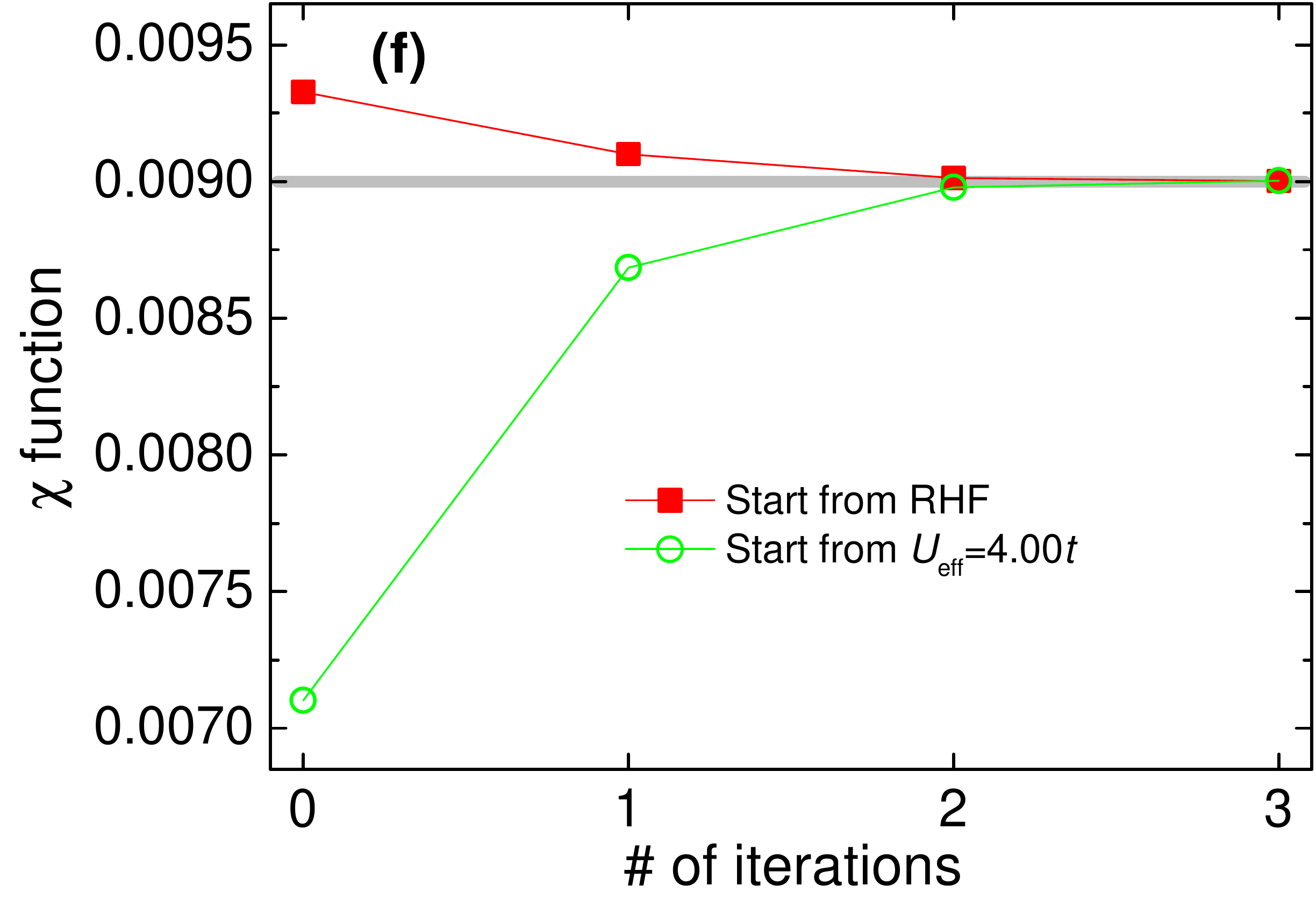}
\caption{\label{fig:SCFUHFTest}(Color online) Convergence of the self-consistent FT-AFQMC procedure. Results are shown for a $16\times4$ system with $U=4t,\beta t=40$ and AFM pinning field $h=0.10t$ applied to both edges in $x$ direction, targeting an overall electron filling $\langle\hat{n}\rangle=0.875$. The left column (a and b) shows results of staggered spin and hole density respectively, starting from the RHF trial Hamiltonian, while the middle column (c and d) presents the corresponding results starting from the UHF trial Hamiltonian with $U_{\text{eff}}=U$. The difference at each stage of the iteration with respect to the final converged results is shown in the insets. The right column shows energy (e), electron filling (inset in e), and $\chi$ (f) versus iteration number. }
\end{figure*}

\begin{figure}[t]
\centering
\includegraphics[width=0.92\columnwidth]{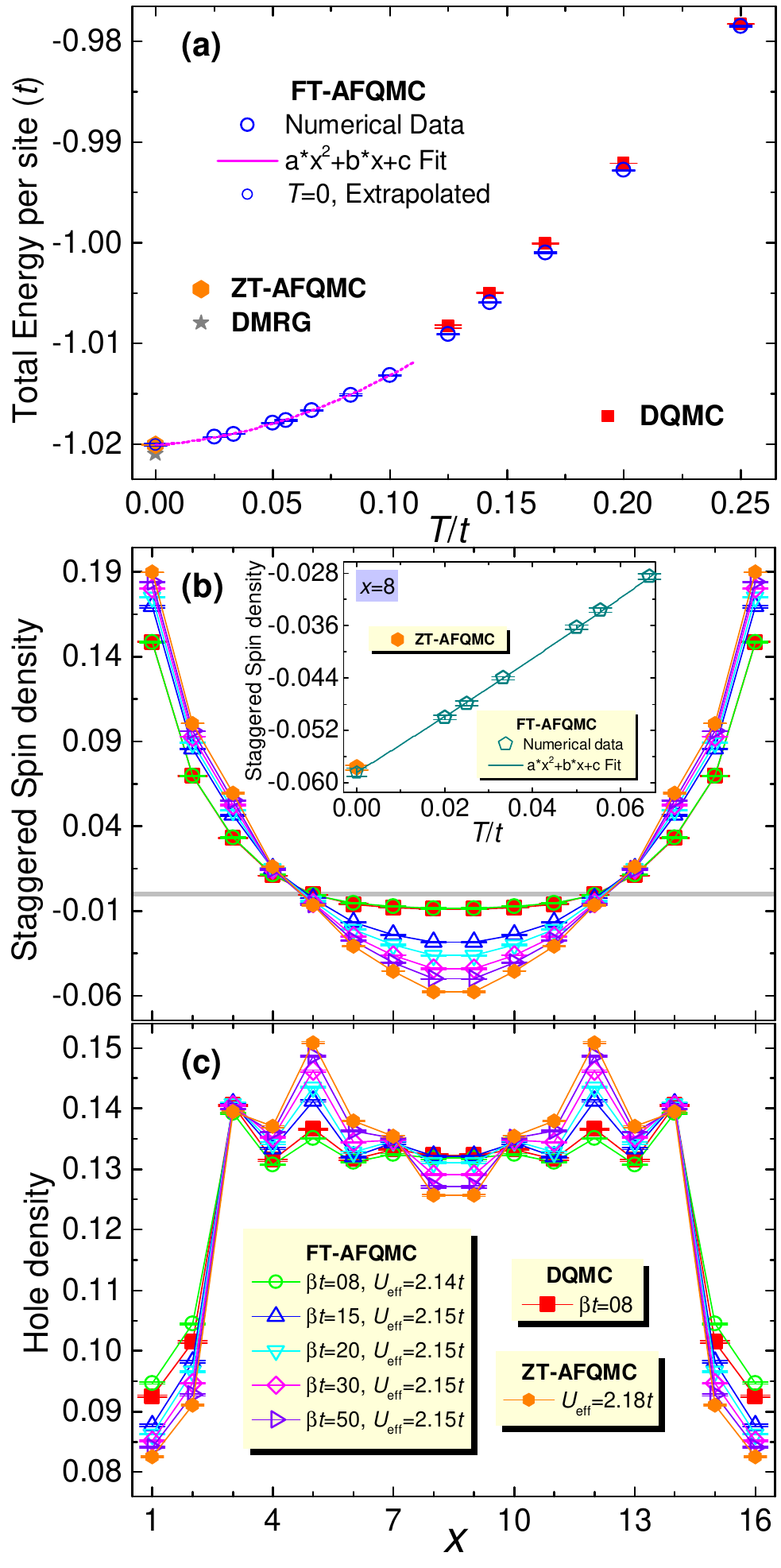}
\caption{\label{fig:SCFUHFU04h010}(Color online) Computed results of the (a) energy, (b) staggered spin density and (c) hole density at a variety of temperatures, after convergence of self-consistency procedure. The system is $16\times4$ with $U=4t$ and $1/8$ hole doping, with AFM pinning fields of $h=0.10t$ applied on both edges along the $y$-direction. The converged $U_{\text{eff}}$ parameter for the trial Hamiltonian is also shown at each $T$. FT-AFQMC results systematically approach those from ZT-AFQMC as temperature is lowered. (b) and (c) share the same symbols, legends and scales of x axis. The inset in (b) presents the staggered spin density at $i_x=8$ and a quadratic fit at low $T$. }
\end{figure}

\subsection{Numerical results from FT-AFQMC calculations with self-consistent procedure}
\label{sec:SCFUHFResults}

The FT-AFQMC results presented in Sec.~\ref{sec:FTCPMCWithRHF} are from one-shot calculations with a specific trial Hamiltonian (namely RHF). In such calculations, there is no internal mechanism to indicate whether the results are optimal or not. The self-consistent procedure developed for $T>0$ in this work serves as such a mechanism. In this part, we illustrate and benchmark the self-consistent approach. As mentioned, we will use a $16\times4$ system at  $\langle\hat{n}\rangle=0.875$ with AFM pinning fields. The local spin and hole densities, $s_i^z=\langle\hat{n}_{i\uparrow}-\hat{n}_{i\downarrow}\rangle/2$ and $n_i^h=1-\langle\hat{n}_{i\uparrow}+\hat{n}_{i\downarrow}\rangle$, will serve as probes for spin and charge order. To plot the spin order, we will often use the staggered spin density $(-1)^{i_x+i_y}s_i^z$. Translational symmetry in $y$- and the mirror symmetry in $x$-direction are applied to reduce statistical errors.

We start by testing the general stability of the self-consistent FT-AFQMC procedure, in Fig.~\ref{fig:SCFUHFTest}. We carry out two separate self-consistent processes, starting the calculation with RHF and UHF trial Hamiltonians, respectively. The UHF trial Hamiltonian is prepared from the UHF mean-field solution of the many-body system at the desired temperature ($\beta t=40$), using original parameters in the many-body Hamiltonian ($U_{\text{eff}}=4t$). The results are shown in panels (a)(b) and (c)(d), respectively. We observe that FT-AFQMC calculations starting from both RHF and UHF trial Hamiltonians converge within four iterations. While it is possible for the self-consistent iteration to get stuck in a local minimum, the tests we have carried out at finite-temperatures resulted in the same converged results for all studied observables, independent of which initial trial Hamiltonian is used. (Interestingly, the $\chi$ function monotonically increases with the $U_{\text{eff}}=4t$  UHF starting trial Hamiltonian, which is different from the behavior in ZT-AFQMC calculations~\cite{MingPu2016b}. This might be from how finite temperature is accounted for in the trial density matrix.)

We now study the temperature dependence of the self-consistent FT-AFQMC procedure. We will carry out calculations for two systems, with $U=4t$ and $U=6t$, at temperatures as low as $\beta t=50$. The lowest temperature accessible with standard DQMC is $\beta t\sim8$ and $5$, with average signs of $0.043$ and $0.016$, respectively. As we shall see, the behaviors of the system at such temperatures are very different from those at low-temperatures and in the ground-state limit, highlighting the fundamental need for FT methods which control the sign problem. The ZT-AFQMC results from self-consistent calculations for these two systems are presented and discussed in Appendix~\ref{sec:SCFZTCPMCResult}.

The computed total energy, staggered spin density and hole density are shown in Fig.~\ref{fig:SCFUHFU04h010} for the $U=4t$ system, after convergence of the self-consistent procedure. The results from DQMC at $\beta t=8$ and DMRG at $T=0$ are also presented as benchmark. The final converged value of the $U_{\text{eff}}$ parameter in the UHF trial Hamiltonian is also indicated for each temperature. As illustrated in Fig.~\ref{fig:SCFUHFU04h010}(a), the results of total energy from AFQMC calculations are highly accurate for this system, with relative systematic error $0.07\%$ at $\beta t=8$ and $0.09\%$ for the ground state. Results of a polynomial fit is shown at low temperatures, which gives an extrapolated answer at $T=0$ fully consistent with ZT-AFQMC. At $\beta t=8$, the FT-AFQMC calculation yields staggered spin density which is statistically indistinguishable from that of DQMC as shown in Fig.~\ref{fig:SCFUHFU04h010}(b). The hole density in Fig.~\ref{fig:SCFUHFU04h010}(c) is also in good agreement but discrepancy from the constraint is visible, especially near the edges; this behavior is consistent with ZT-AFQMC calculations~\cite{MingPu2016b}. Both results converge monotonically toward the ZT-AFQMC results as temperature is lowered. As shown more quantitatively in the inset of Fig.~\ref{fig:SCFUHFU04h010}(b), the staggered spin density can be fitted very well by a polynomial, and the extrapolated result at $T=0$ is in excellent agreement with the self-consistent ZT-AFQMC result. The self-consistent $U_{\text{eff}}$ parameters shown in Fig.~\ref{fig:SCFUHFU04h010}(b) and (c) only vary slightly as $U_{\text{eff}}/t=2.14\sim2.15$ within a large range of temperature.

\begin{figure}[t]
\centering
\includegraphics[width=0.93\columnwidth]{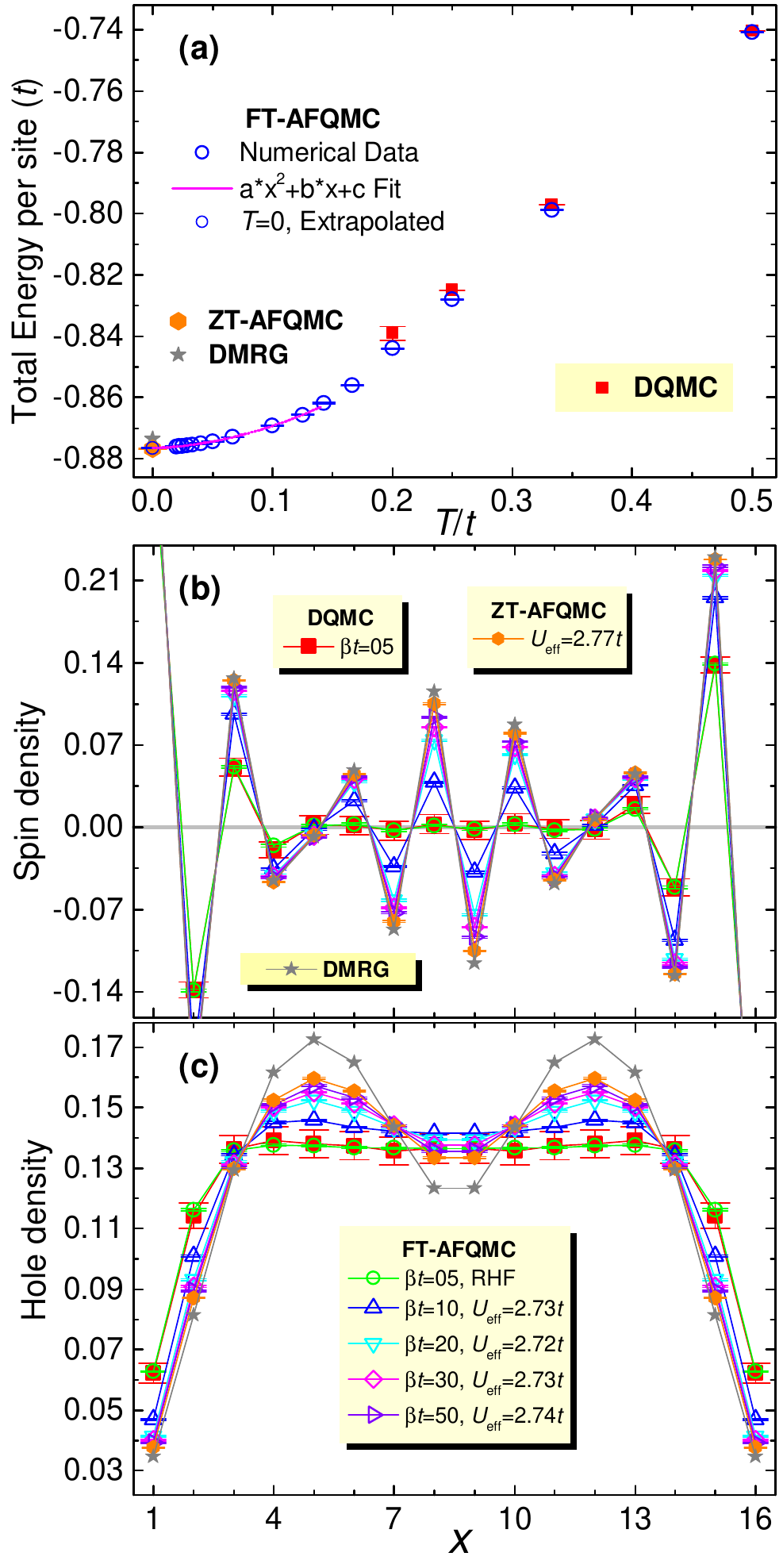}
\caption{\label{fig:SCFGSUHFU06h025All}(Color online) Development of antiferromagnetic correlations and charge order as temperature is lowered. The system is $16\times4$, $U=6t$ and $1/8$ hole doping, with AFM pinning fields of $h=0.10t$ applied on both edges along the $y$-direction. The AFQMC results are from self-consistent calculations. Energies are shown in (a), and spin density and hole density are plotted in panels (b) and (c), which share the same symbols, legends and scales of the x axis. }
\end{figure}

We next turn to physically more relevant and challenging system with $U=6t$. Many of the characteristics seen in the benchmark in the $U=4t$ systems are also seen here. In Fig.~\ref{fig:SCFGSUHFU06h025All}, results are shown for the total energy, the spin order and hole density for a temperature range $\beta=4$ to $50$. The sign problem prevents DQMC calculations from reaching much below $\beta=5$, indicated by the large error bars of the results in all three panels in Fig.~\ref{fig:SCFGSUHFU06h025All}. This is too high a temperature for the antiferromagnetic correlations to develop, as seen from the results at lower temperatures. At $\beta t=5$, the pinning fields near the edge introduce short-range antiferromagnetic correlations. The hole density in the system also responds but remains flat outside of the edge region. As temperature is lowered, the AFM order systematically increases in the self-consistent FT-AFQMC results. The hole density starts to exhibit inhomogeneities, and develop peaks near the nodal lines where the AFM order shows a phase change. The spin and hole densities approach the ZT-AFQMC results but with visible difference even at $\beta t=20$. The results from FT-AFQMC at $\beta t=50$ and ZT-AFQMC are virtually indistinguishable. Compared to exact ground-state DMRG results, the spin density from AFQMC is highly accurate, while the hole density shows noticeable discrepancy. This is consistent with the observation from ground-state studies~\cite{MingPu2016b} where more systems were tested. As shown in Fig.~\ref{fig:SCFGSUHFU06h025All}(a), the total energy per site is also exact at high temperatures, compared with DQMC results. At lower temperatures, the computed energies extrapolate to a value of $-0.8765(3)$, consistent with the ZT-AFQMC results of $-0.8768(2)$, which has a relative error of about $0.36\%$ compared to the exact ground-state energy of $-0.8735$ from DMRG.

Besides obtaining the results of thermodynamic properties, the self-consistent procedure in FT-AFQMC calculations also determines an effective interaction strength $U_{\text{eff}}$. The computed spin and charge densities from the IP Hamiltonian in Eq.~(\ref{eq:IPHamiltonian}) actually provide a reasonable approximation to the many-body results. As expected, the optimum $U_{\text{eff}}$ increases with interaction strength $U$ in the many-body Hamiltonian. Similar to the $U=4t$ case, the $U_{\text{eff}}$ at $T>0$ is found to be smaller than that at $T=0$ at $U=6t$, although $U_{\text{eff}}$ only varies slightly in a wide range of low temperatures.

\section{Discussion and Summary}
\label{sec:SumDiscuss}

The self-consistent FT-AFQMC method developed in this work allows finite-temperature QMC computations to systematically improve the accuracy while removing the exponential computational scaling from the sign/phase problem. We have studied the form of the trial Hamiltonian or trial density matrix. Compared to ground-state calculations, for which the constraint is defined in terms of the ground-state  trial wave function, the FT-AFQMC formalism involves accounting for the temperature. The self-consistent constraint we have discussed aims to generate a ${\hat B_T}$ by optimizing $e^{-\beta {\hat H_T}}$ with respect to the many-body calculation at the desired temperature $\beta$. The alternative we also tested, of using a ground-state ${\hat H_T}$, takes the view of optimizing an effective {\it Hamiltonian}. No significant difference was observed in the performance of these in the Hubbard model. Another approach which will be interesting to test is to, as random walk proceeds (varying $\ell$), refine the constraint $e^{-\tau {\hat H_T}}$ ``dynamically'' with $\tau$, which corresponds to $\prod_n^{M-\ell} {\hat B_T}$ in the constraining conditions in Eq.~(\ref{eq:PartConstrctPathTrialBT}), as the random walk proceeds (varying $\ell$). Applying the computed single-particle density matrix as constraint to form a self-consistent procedure is also possible, as discussed in Ref.~\onlinecite{MingPu2016b} for the ground state. We will leave these to a future study.

As mentioned, the self-consistent approach produces a rigorous procedure to identify an optimal independent-particle treatment of the many-body system. In the Hubbard model, this takes the form of an effective interaction. In a more general context, one can imagine this procedure as a way to optimize an exchange-correlation functional within the context of density-functional theory (DFT). The optimized parametrization of a particular DFT flavor can then be applied to larger systems or other materials. Similarly, one could imagine using the approach to help determine a ``$U$'' parameter for DFT+$U$ type of calculations in correlated materials.

The finite-temperature extension of AFQMC allows many potential applications. Among these are direct calculations to study temperature dependence and obtain thermodynamic information in correlated electron models and materials. It could also potentially be combined with embedding methods such as dynamical mean-field theory~\cite{Georges1996}, acting as an impurity solver.

To summarize, we have presented a self-consistent finite-temperature AFQMC approach to study many-fermion systems. The method shares the basic formalism of DQMC, but controls the sign or phase problem with a constraint. The calculations are formulated as branching random walks with importance sampling. The constraint is applied with a trial Hamiltonian in  an independent-particle form, or a density matrix, and is optimized by a self-consistent feedback from the AFQMC calculation. The approach complements the ZT-AFQMC which has been widely applied in lattice models and in realistic systems in condensed matter and quantum chemistry. In this paper, we presented the finite-temperature algorithm in detail, including many technical and numerical aspects, which will help make the implementation of FT-AFQMC method to other correlated fermion systems more straightforward. With the new method and with improved numerical stabilization procedures (further described in Appendix~\ref{sec:SCFZTCPMCResult}), calculations can reach very low temperatures (as low as $1/80$ in the Hubbard model).

In addition to presenting the method, we also carried out a careful benchmark, and studied its behavior in detail. Different forms of the trial Hamiltonian were tested. Comparisons were made with DQMC where the latter could be performed, as well as with ZT-AFQMC and ED and DMRG in the ground state. The FT-AFQMC method is indistinguishable from DQMC at high temperatures, and connects smoothly to the ZT-AFQMC at very low temperatures. The benchmark results  show that the accuracy of FT-AFQMC method for thermodynamic properties is comparable to or even better than that of ZT-AFQMC method for ground-state properties. The results in the Hubbard model demonstrate that the approach opens up regimes in temperature where new physics takes place and which were previously inaccessible by DQMC. It is hoped that applications to study thermodynamic properties in a variety of correlated fermion models will now be possible with developments in this work.

\begin{acknowledgments}

Y.Y.H acknowledges H. Yao and F. Wang for valuable discussions. Y.Y.H. sincerely thanks the hospitality of school of physics, Northwest University, where part of this work was done. Y.Y.H and Z.Y.L acknowledge fundings from the National Science Foundation of China under Grant Nos. 11774422, 11474356. Y.Y.H is also supported by the Outstanding Innovative Talents Cultivation Funded Programs 2016 of Renmin University of China. S.Z.~acknowledges support from NSF (Grant No. DMR-1409510). We thank the Physical Laboratory of High Performance Computing in Renmin University of China, the computational facilities at College of William and Mary, and the Tianhe-2A platform at the National Supercomputer Center in Guangzhou for their technical support and generous allocation of CPU time. The Flatiron Institute is a division of the Simons Foundation.
\end{acknowledgments}

\appendix

\section{Some further details in FT-AFQMC method}
\label{sec:FTCPMCOtherDetail}

In this appendix, we further discuss several technical details of the FT-AFQMC implementation.

In Sec.~\ref{sec:ImportSample}, we mentioned that there are two somewhat different ways to implement FT-AFQMC depending on the connection between $\hat{H}_T$ and $\hat{H}_0$. If $\hat{H}_T$ is closely related to the free fermion part $\hat{H}_0$, e.g., with the RHF or UHF form, we need to insert the $e^{-\Delta\tau(\hat{H}_I+\hat{H}_0-\hat{H}_T)}=\sum_{\mathbf{x}} p(\mathbf{x})\hat{B}_I(\mathbf{x})$ operator at every time slice. In this case we have, for the $\ell$-th time slice
\begin{equation}
\label{eq:RatioNew}
\begin{split}
&\frac{\mathcal{P}_\ell^T}{\mathcal{P}_{\ell-1}^T} = p(\mathbf{x}_\ell) \frac{\text{Tr}\Big[\Big(\prod\limits_{n=1}^{M-\ell}\hat{B}_T\Big){\color{blue}{\hat{B}_I(\mathbf{x}_\ell)}}\hat{B}_T
\Big(\prod\limits_{m=1}^{\ell-1}\hat{B}_m\Big)\Big]}
{\text{Tr}\Big[\Big(\prod\limits_{n=1}^{M-\ell}\hat{B}_T\Big)\hat{B}_T\Big(\prod\limits_{m=1}^{\ell-1}\hat{B}_m\Big)\Big]} \\
&=p(\mathbf{x}_\ell)\frac{\text{det}\Big[\mathbf{I}_{2N_s}+\Big(\prod\limits_{n=1}^{M-\ell}\mathbf{B}_T\Big)
{\color{blue}{\mathbf{B}_I(\mathbf{x}_\ell)}}\mathbf{B}_T\Big(\prod\limits_{m=1}^{\ell-1}\mathbf{B}_{m}\Big)\Big]}
{\text{det}\Big[\mathbf{I}_{2N_s}+\Big(\prod\limits_{n=1}^{M-\ell}\mathbf{B}_T\Big)\mathbf{B}_T
\Big(\prod\limits_{m=1}^{\ell-1}\mathbf{B}_{m}\Big)\Big]}\,,
\end{split}
\end{equation}
where the notations are the same as the main text. In the second way, the trial Hamiltonian $\hat{H}_T$ is assumed to be quite different from $\hat{H}_0$, and we need to replace $\hat{B}_T$ by $\hat{B}=\hat{B}_I(\mathbf{x})\hat{B}_K$ for every time slice. In this case, we first replace $\hat{B}_T$ by $\hat{B}_K$ and then insert the $e^{-\Delta\tau\hat{H}_I}=\sum_{\mathbf{x}} p(\mathbf{x})\hat{B}_I(\mathbf{x})$ operator to evaluate $\mathcal{P}_\ell^T/\mathcal{P}_{\ell-1}^T$ during the random walk:
\onecolumngrid \begin{equation}
\label{eq:RatioNew011}
\begin{split}
\frac{\mathcal{P}_\ell^T}{\mathcal{P}_{\ell-1}^T} &= p(\mathbf{x}_\ell) \frac{\text{Tr}\Big[\Big(\prod_{n=1}^{M-\ell}\hat{B}_T\Big){\color{blue}{\hat{B}_I(\mathbf{x}_\ell)\hat{B}_K}}\hat{B}_{\ell-1}\cdots\hat{B}_2\hat{B}_1\Big]}
{\text{Tr}\Big[\Big(\prod_{n=1}^{M-\ell}\hat{B}_T\Big)\hat{B}_T\hat{B}_{\ell-1}\cdots\hat{B}_2\hat{B}_1\Big]} \\
&= p(\mathbf{x}_\ell) \frac{\text{Tr}\Big[\Big(\prod_{n=1}^{M-\ell}\hat{B}_T\Big){\color{blue}{\hat{B}_I(\mathbf{x}_\ell)}}\hat{B}_K\hat{B}_{\ell-1}\cdots\hat{B}_1\Big]}
{\text{Tr}\Big[\Big(\prod_{n=1}^{M-\ell}\hat{B}_T\Big)\hat{B}_K\hat{B}_{\ell-1}\cdots\hat{B}_1\Big]}\cdot
\frac{\text{Tr}\Big[\Big(\prod_{n=1}^{M-\ell}\hat{B}_T\Big){\color{magenta}{\hat{B}_K}}\hat{B}_{\ell-1}\cdots\hat{B}_1\Big]}
{\text{Tr}\Big[\Big(\prod_{n=1}^{M-\ell}\hat{B}_T\Big){\color{magenta}{\hat{B}_T}}\hat{B}_{\ell-1}\cdots\hat{B}_1\Big]} \\
&= p(\mathbf{x}_\ell)\frac{\text{det}\Big[\mathbf{I}_{2N_s}+\Big(\prod_{n=1}^{M-\ell}\mathbf{B}_T\Big){\color{blue}{\mathbf{B}_I(\mathbf{x}_\ell)}}\mathbf{B}_K\mathbf{B}_{\ell-1}\cdots\mathbf{B}_1\Big]}
{\text{det}\Big[\mathbf{I}_{2N_s}+\Big(\prod_{n=1}^{M-\ell}\mathbf{B}_T\Big)\mathbf{B}_K\mathbf{B}_{\ell-1}\cdots\mathbf{B}_1\Big]}\cdot
\frac{\text{det}\Big[\mathbf{I}_{2N_s}+\Big(\prod_{n=1}^{M-\ell}\mathbf{B}_T\Big){\color{magenta}{\mathbf{B}_K}}\mathbf{B}_{\ell-1}\cdots\mathbf{B}_1\Big]}
{\text{det}\Big[\mathbf{I}_{2N_s}+\Big(\prod_{n=1}^{M-\ell}\mathbf{B}_T\Big){\color{magenta}{\mathbf{B}_T}}\mathbf{B}_{\ell-1}\cdots\mathbf{B}_1\Big]},
\end{split}
\end{equation} \twocolumngrid
\noindent{where the second and first ratios in the last line correspond to the insertion of $\mathbf{B}_I(\mathbf{x}_\ell)$ matrix and replacement of $\mathbf{B}_T$ by $\mathbf{B}_K$ matrix}. To evaluate $\mathcal{P}_\ell^T/\mathcal{P}_{\ell-1}^T$ in Eq.~(\ref{eq:RatioNew011}) numerically, we need to calculate the determinant $\text{det}[\mathbf{I}_{2N_s}+\big(\prod_{n=1}^{M-\ell}\mathbf{B}_T\big)\mathbf{B}_K\mathbf{B}_{\ell-1}\cdots\mathbf{B}_1]$ and the static single-particle Green's function at every time slice, which results in extra computational cost compared to the first way in Eq.~(\ref{eq:RatioNew}).

We next describe our numerical stabilization procedures. In DMQC and FT-AFQMC method, the numerical stablization procedures are similar and they both involves the static single-particle Green's function matrix $\mathbf{G}(\tau,\tau)=\{G_{i\sigma,j\sigma^\prime}=\langle c_{i\sigma}c_{j\sigma^\prime}\rangle_{\tau}\}$ at $\tau=\ell\Delta\tau$. Generally, its matrix element takes the form
\begin{equation}
\begin{split}
\label{eq:SinglePart011}
G_{i\sigma,j\sigma^\prime} &= \frac{ \text{Tr}[ \hat{B}(\beta,\tau) c_{i\sigma}c^+_{j\sigma^\prime} \hat{B}(\tau, 0) ] }
{ \text{Tr}[ \hat{B}(\beta,\tau) \hat{B}(\tau, 0) ] }  \\
&= \Big\{\big[ \mathbf{I}_{2N_s} + \mathbf{B}(\tau,0)\mathbf{B}(\beta,\tau) \big]^{-1}\Big\}_{i\sigma,j\sigma^\prime}.
\end{split}
\end{equation}
For DQMC, $\hat{B}(\beta,\tau)=\hat{B}_M\cdots\hat{B}_{\ell+2}\hat{B}_{\ell+1}$ and $\hat{B}(\tau, 0)=\hat{B}_{\ell}\cdots\hat{B}_2\hat{B}_1$, with $\mathbf{B}(\beta,\tau)=\mathbf{B}_M\cdots\mathbf{B}_{\ell+2}\mathbf{B}_{\ell+1}$ and $\mathbf{B}(\tau, 0)=\mathbf{B}_{\ell}\cdots\mathbf{B}_2\mathbf{B}_1$. For FT-AFQMC algorithm, $\hat{B}(\beta,\tau)=\prod_{n=1}^{M-\ell}\hat{B}_T$ and $\hat{B}(\tau, 0)=\hat{B}_{\ell}\cdots\hat{B}_2\hat{B}_1$, with $\mathbf{B}(\beta,\tau)=\prod_{n=1}^{M-\ell}\mathbf{B}_T$, $\mathbf{B}(\tau, 0)=\mathbf{B}_{\ell}\cdots\mathbf{B}_2\mathbf{B}_1$, instead.

We apply the column-pivoted QR algorithm~\cite{LOH2005,Tomas2012} to perform the following decompositions for $\mathbf{B}(\beta,\tau)$ and $\mathbf{B}(\tau, 0)$ matrices:  $\mathbf{B}(\beta,\tau)=\mathbf{V}_L\mathbf{D}_L\mathbf{U}_L$ and $\mathbf{B}(\tau, 0)=\mathbf{U}_R\mathbf{D}_R\mathbf{V}_R$, where $\mathbf{U}_R,\mathbf{U}_L,\mathbf{V}_R,\mathbf{V}_L$ are $2N_s\times 2N_s$ matrices and $\mathbf{U}_R,\mathbf{U}_L$ are unitary, and $\mathbf{D}_R,\mathbf{D}_L$ are $2N_s\times 2N_s$ real diagonal matrices. To calculate the Green's function matrix in Eq.~(\ref{eq:SinglePart011}), we further separate $\mathbf{D}_R$ as $\mathbf{D}_\text{R}^{\text{max}}\mathbf{D}_\text{R}^{\text{min}}$, which are both $2N_s\times 2N_s$ real diagonal matrices satisfying the following condition: if $|(\mathbf{D}_\text{R})_{ii}|\ge1$, then $(\mathbf{D}_\text{R}^{\text{max}})_{ii}=|(\mathbf{D}_\text{R})_{ii}|$ and $(\mathbf{D}_\text{R}^{\text{min}})_{ii}=\text{Sgn}[(\mathbf{D}_\text{R})_{ii}]$; if $|(\mathbf{D}_\text{R})_{ii}|<1$, then $(\mathbf{D}_\text{R}^{\text{max}})_{ii}=1$ and $(\mathbf{D}_\text{R}^{\text{min}})_{ii}=(\mathbf{D}_\text{R})_{ii}$.
Similarly, we have $\mathbf{D}_\text{L}=\mathbf{D}_\text{L}^{\text{min}}\mathbf{D}_\text{R}^{\text{max}}$ following the same definitions. We can the ncalculate the single-particle Green's function matrix in Eq.~(\ref{eq:SinglePart011}) in a numerically stable manner~\cite{LOH2005,Tomas2012} as
\onecolumngrid \begin{equation}
\begin{split}
\label{eq:FTStable002}
\mathbf{G}(\tau,\tau)
&=  (\mathbf{U}_\text{L})^{-1} \Big[(\mathbf{U}_\text{R})^{-1} (\mathbf{U}_\text{L})^{-1} +  \mathbf{D}_\text{R}\mathbf{V}_\text{R}
     \mathbf{V}_\text{L}\mathbf{D}_\text{L} \Big]^{-1} (\mathbf{U}_\text{R})^{-1} \\
&= (\mathbf{U}_\text{L})^{-1} \Big[(\mathbf{U}_\text{L}\mathbf{U}_\text{R})^{-1} +  \mathbf{D}_\text{R}^{\text{max}}\mathbf{D}_\text{R}^{\text{min}}\mathbf{V}_\text{R}
     \mathbf{V}_\text{L}\mathbf{D}_\text{L}^{\text{min}}\mathbf{D}_\text{L}^{\text{max}} \Big]^{-1} (\mathbf{U}_\text{R})^{-1} \\
&= (\mathbf{U}_\text{L})^{-1} (\mathbf{D}_\text{L}^{\text{max}})^{-1}\Big[(\mathbf{D}_\text{R}^{\text{max}})^{-1}
(\mathbf{U}_\text{L}\mathbf{U}_\text{R})^{-1}(\mathbf{D}_\text{L}^{\text{max}})^{-1}  +  \mathbf{D}_\text{R}^{\text{min}}\mathbf{V}_\text{R}
     \mathbf{V}_\text{L}\mathbf{D}_\text{L}^{\text{min}} \Big]^{-1} (\mathbf{D}_\text{R}^{\text{max}})^{-1} (\mathbf{U}_\text{R})^{-1}.
\end{split}
\end{equation} \twocolumngrid
From Eq.~(\ref{eq:FTStable002}), we can observe that the matrix $[(\mathbf{D}_\text{R}^{\text{max}})^{-1}
(\mathbf{U}_\text{L}\mathbf{U}_\text{R})^{-1}(\mathbf{D}_\text{L}^{\text{max}})^{-1} + \mathbf{D}_\text{R}^{\text{min}}\mathbf{V}_\text{R}
\mathbf{V}_\text{L}\mathbf{D}_\text{L}^{\text{min}}]$, whose inverse matrix needs to be calculated, should have all its matrix elements around $-10^0\sim10^0$. For time-displaced single-particle Green's function matrices $\mathbf{G}(\tau,0)$ and $\mathbf{G}(0, \tau)$, there are similar formulas to calculate them in a numerically stable manner. Practically, the numerical stablization is carried out with a suitably chosen number of imaginary time slice as interval.

\section{Numerical results from self-consistent ZT-AFQMC calculations}
\label{sec:SCFZTCPMCResult}

\begin{figure*}
\centering
\includegraphics[width=0.671\columnwidth]{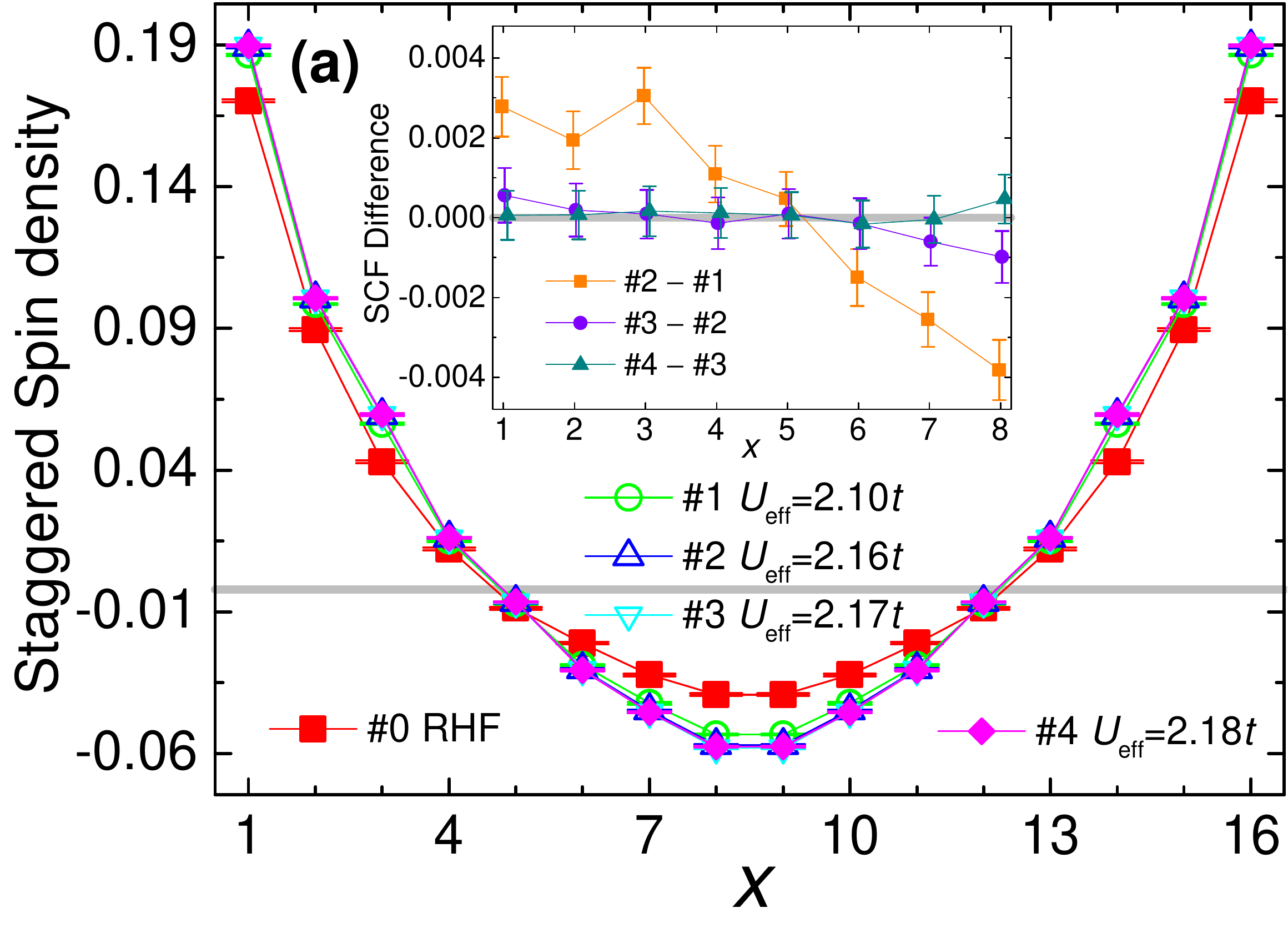}
\includegraphics[width=0.663\columnwidth]{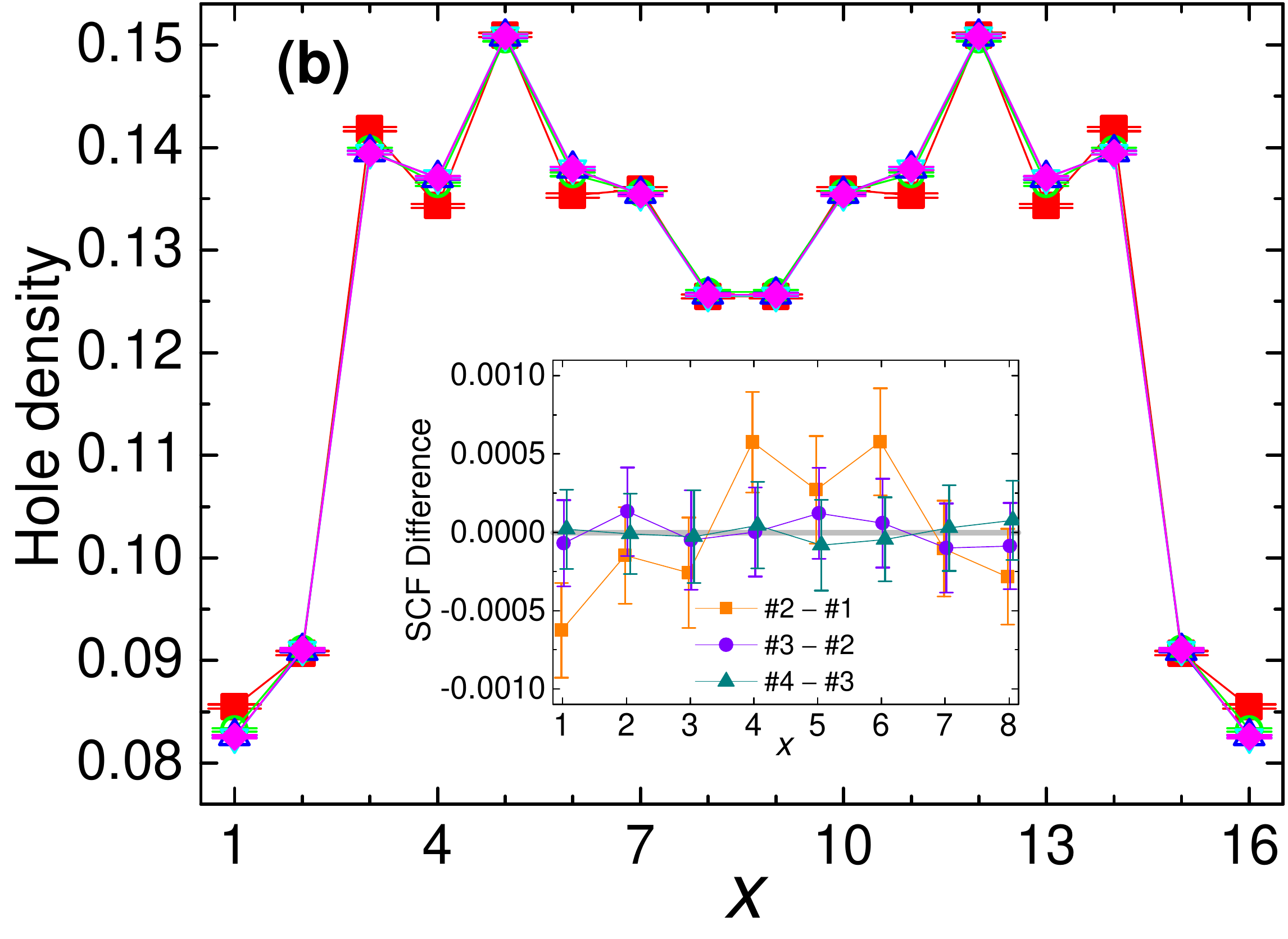}
\includegraphics[width=0.700\columnwidth]{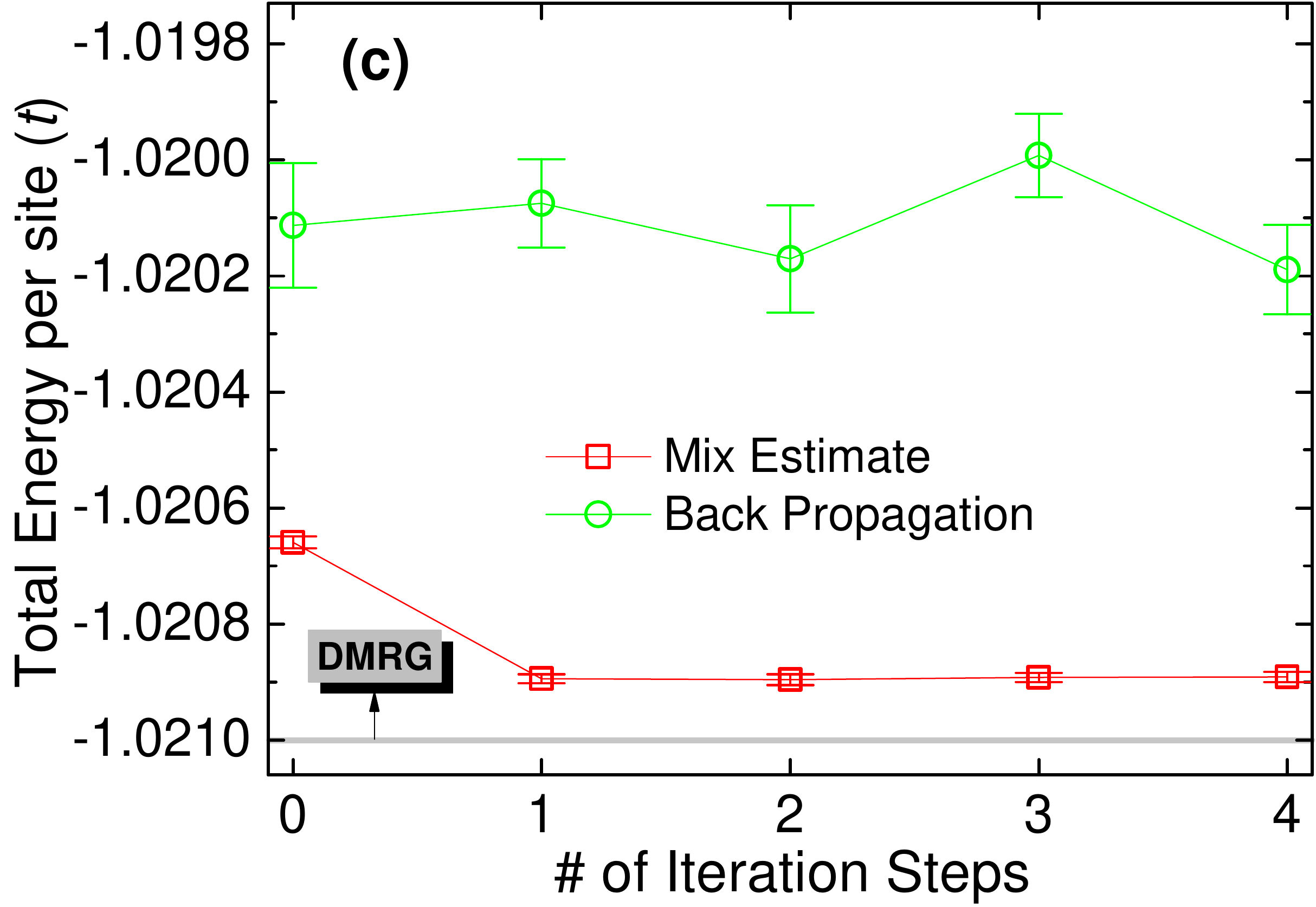} \\
\includegraphics[width=0.671\columnwidth]{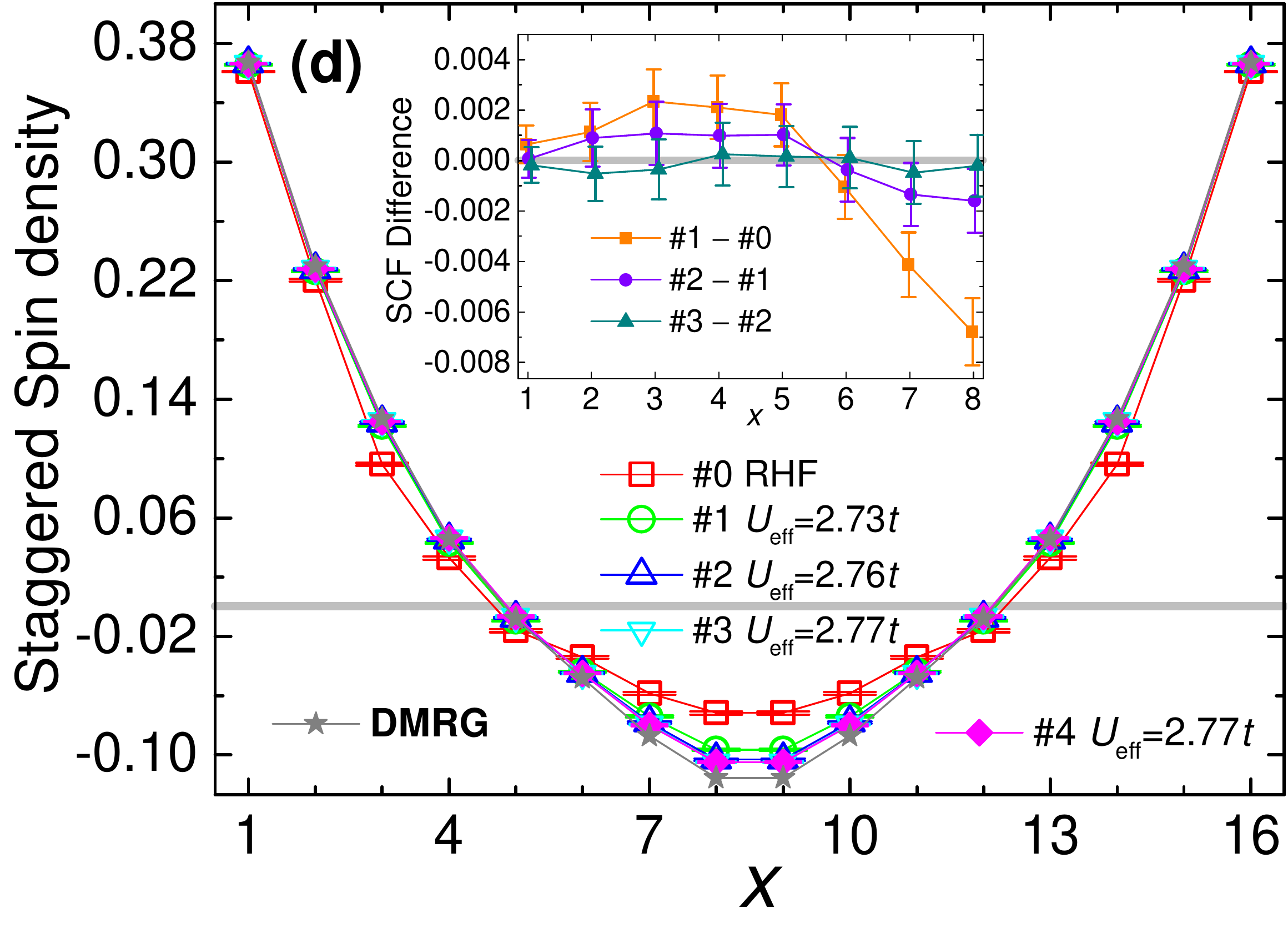}
\includegraphics[width=0.663\columnwidth]{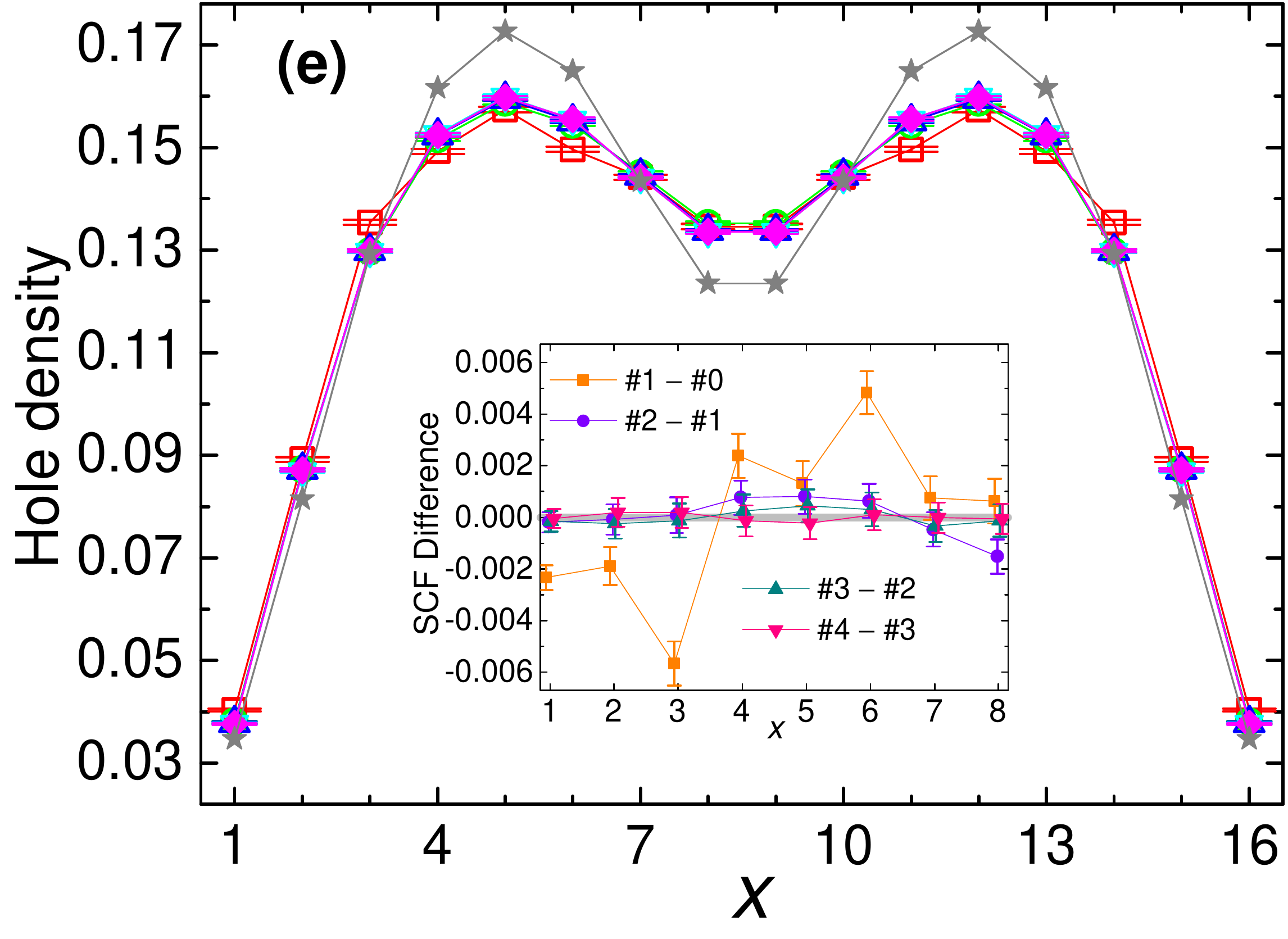}
\includegraphics[width=0.700\columnwidth]{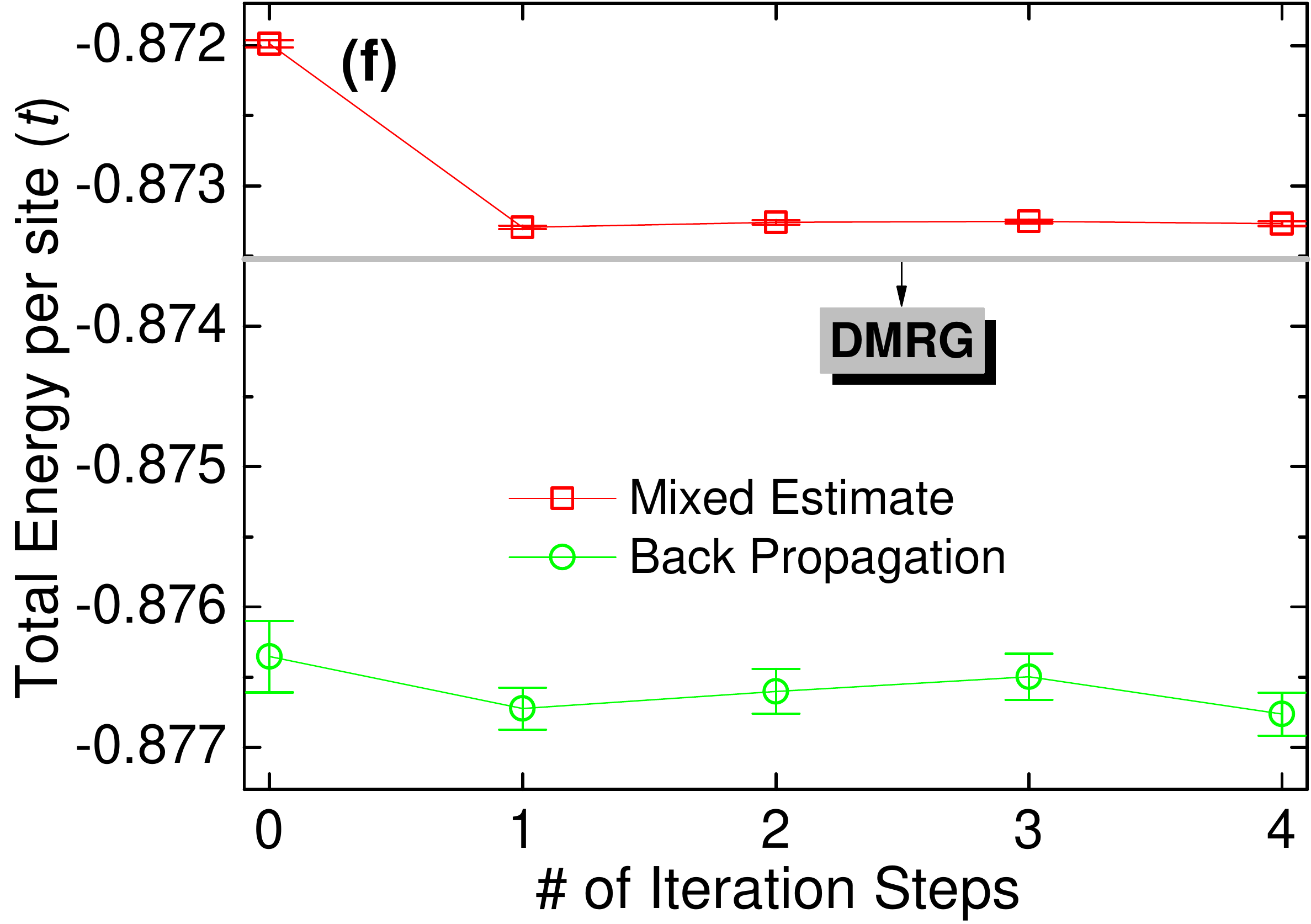}
\caption{\label{fig:ZTCPMCU04U06}(Color online) Numerical results from self-consistent ZT-AFQMC calculations at $T=0$ for $16\times4$ system with $U=4t$ and $U=6t$, with $1/8$ hole doping. AFM pinning fields are applied on both edges along $y$ direction, with $h=0.10t$ for $U=4t$ and $h=0.25t$ for $U=6t$. (a), (b) and (c) are Results of staggered spin density, hole density and total energy per site are plotted in three panels in each row, with the top row for the $U=4t$ system and the bottom row (d)(e)(f) for $U=6t$. (a)(b) and (d)(e) share the same symbols and legends, respectively. The insets in the first two columns show the difference between successive iterations. }
\end{figure*}

In Sec.~\ref{sec:SCFUHFResults}, we presented $T=0$ results from ZT-AFQMC calculations for comparisons with the self-consistent FT-AFQMC results on $16\times4$ systems at $1/8$ hole doping. Here we include the  ZT-AFQMC  results for completeness, summarized in Fig.~\ref{fig:ZTCPMCU04U06}.

The self-consistent calculations started from free electron trial wavefunctions. For both systems, five steps of iterations reached the converged results for the densities, while the total energy  converged within two steps. The converged effective interaction strengths are $U_{\text{eff}}=2.18t$ and $U_{\text{eff}}=2.77t$ for those two systems, respectively. The convergence of staggered spin density and hole density are illustrated by the differences of results between successive iterations as shown in Fig.~\ref{fig:ZTCPMCU04U06}(a)(b) for $U=4t,h=0.10t$ system and Fig.~\ref{fig:ZTCPMCU04U06}(d)(e) for $U=6t,h=0.25t$ system. For the $U=6t$ system, the converged results of staggered spin density is very close to the DMRG results, highlighting the high accuracy. The hole density still has some deviations, similar to the results shown in Ref.~\onlinecite{MingPu2016b}. The results of total energy per site from both mixed estimate and back propagation measurements are shown in the last column. As expected, the error bars of results of total energy per site from mixed estimate measurements is much smaller than those from back propagation measurements as shown in (c)(f). For $U=6t$, the converged results of total energy per site from those two different ways of measurement have relative errors of $0.028\%$ and $0.355\%$ from DMRG, respectively.

\bibliography{FTCPMC_Benchmark}

\end{document}